\documentclass[superscriptaddress,nobibnotes,amsmath,amssymb,notitlepage,twocolumn,prb,longbibliography]{revtex4-1}

\usepackage{bm,mathptmx,braket}
\usepackage{amsmath}
\usepackage{physics}
\usepackage{graphicx,color,hyperref}
\usepackage[caption=false]{subfig}
\hypersetup{colorlinks=true, linkcolor=blue, citecolor=blue, urlcolor=blue} 

\graphicspath{{./img/}}

\newcommand{\beq}[1]{\begin{equation}\label{#1}}
\newcommand{\eep}{\;.\end{equation}}
\newcommand{\eec}{\;,\end{equation}}
\newcommand{\eeq}{\end{equation}}

\newcommand{\lb}{\left(}
\newcommand{\rb}{\right)}

\renewcommand{\a}{\alpha}

\renewcommand{\d}{\delta}
\newcommand{\ep}{\epsilon}

\renewcommand{\th}{\theta}
\newcommand{\s}{\sigma}

\newcommand{\om}{\omega}

\newcommand{\G}{\Gamma}

\newcommand{\Om}{\Omega}
\newcommand{\ra}{\rightarrow}

\DeclareMathAlphabet{\mathcal}{OMS}{cmsy}{m}{n} 

\newcommand{\Df}{\mathcal{D}}    

\newcommand{\bigO}{O} 

\renewcommand{\vec}[1]{{\bf #1}}

\newcommand{\kv}{\vec{k}}

\newcommand{\geo}{\mathrm{geo}}
\newcommand{\sgn}{\mathrm{sgn}}

\begin{document}

\title{Orbital magnetization reveals multiband topology}


\newcommand{\TCM}{{Theory of Condensed Matter Group, Cavendish Laboratory, University of Cambridge, J.\,J.\,Thomson Avenue, Cambridge CB3 0HE, UK}}
\newcommand{\UoM}{Department of Physics and Astronomy, University of Manchester, Oxford Road, Manchester M13 9PL, UK}
\newcommand{\KITP}{Kavli Institute for Theoretical Physics, University of California, Santa Barbara, CA 93106, USA}


\author{Chun Wang Chau}
\email{cwc61@cam.ac.uk}
\affiliation{\TCM}

\author{Robert-Jan Slager}
\email{robert-jan.slager@manchester.ac.uk}
\affiliation{\UoM}
\affiliation{\TCM}

\author{Wojciech J. Jankowski}
\email{wjj25@cam.ac.uk}
\affiliation{\TCM}
\affiliation{\KITP}

\date{\today}

\begin{abstract}
    We demonstrate that nontrivial multiband topological invariants of electronic wavefunctions can be revealed through orbital magnetization responses to external magnetic fields. We find that decomposing orbital magnetization into energetic and quantum-geometric contributions allows one to deduce nontrivial multiband \mbox{topology}, provided knowledge of the energy spectrum. We showcase our findings in general effective models with multiband Euler topology. We moreover identify such multiband topological invariants in effective models of strontium ruthenate ($\text{Sr}_2 \text{Ru} \text{O}_4$), which may in principle be verified in the state-of-the-art doping-dependent magnetization measurements. Our reconstruction scheme for multiband invariants sheds a~topological perspective on the multiorbital effects in materials realizing unconventional phenomenologies of orbital currents or multiband superconductivity. 
\end{abstract}

\maketitle

\section{Introduction} Multistate effects play an important role in physics, from nonlinear optics~\cite{Boyd2019, Jankowski2024PRL, Avdoshkin2025, Jankowski2025PRL}, to diverse correlated quantum phenomena such as superconductivity~\cite{Peotta2015, Xie2020, Peri2021, Herzog2022a, Gassner2024, Chen2024, Chau2025} and magnetism~\cite{Kang2024, Oh2024, Hu2025, Oh2025}. In condensed matter systems, multistate effects are fundamentally encoded in the combinations of matrix elements of quantum operators between multiple single-particle band eigenstates~\cite{Ahn2020, Ahn2021, Bouhon2023, Yu2024, Jankowski2024PRL, Jankowski2025PRBoptical, Avdoshkin2025, Jain2025}. The underlying multiband geometry~\cite{Bouhon2023, Torma2023, Jain2025} arises from the fact that any quantum operator is a derivative in a conjugate quantum variable~\cite{Ahn2019, Shinada2025}, which translates geometric tensors and differential structures of bands into quantum responses manifested in numerous experiments~\cite{Lai2021, Yue2022, Wang2023a, Gao2023}. Combinations of quantum states can realize not only nontrivial Hilbert-space geometry~\cite{Ahn2021, Bouhon2023, Yu2024}, but also nontrivial multiband topologies that are irreducible to sums of single-band invariants such as Chern numbers in anomalous quantum Hall phases~\cite{Niu1985, Haldane1988, Yu2010} or topological crystalline invariants~\cite{Fu2011, Slager2012, Shiozaki2014, Kruthoff2017, Po2017, Bradlyn2017}.

Topologies associated with finite multiband invariants and multigap conditions~\cite{Zhao2017, Wu2019, Ahn2019, Bouhon2019nonabelian, Bouhon2020, Unal2020, Guo2021,Jiang2021a, Slager2024, Wahl2025} are an increasingly studied topic of interest, yet experimental probes of these exotic topologies are rather scarce. The connection to multiband geometry is of key importance for potential multiband invariant reconstruction schemes. For instance, nonlinear optical responses involving multiband torsion tensors allow one to extract unconventional topological invariants beyond the tenfold classification~\cite{Kitaev2009a, Davoyan2024, Jankowski2024PRL, Jankowski2024PRBHopf, Jankowski2025gerbe}. A longstanding challenge includes finding electromagnetic probes for multiband topological Euler invariants~\cite{Jankowski2025PRBoptical, Jain2025, Chau2025} in electronic materials, which at the time of writing, lack a general measurement scheme independent of band energies. In this context, a~particularly underexplored and promising direction entails potential novel responses to magnetic fields beyond the recently studied geometric responses of Euler bands to electric fields~\cite{Jankowski2025PRBoptical, Jain2025}. Magnetic field probes are of particularly high promise for revealing nontrivial multiband topologies, as these (i) reflect the exotic Landau levels arising from multiband topologies~\cite{Bouhon2019nonabelian, Herzog2022, Guan2022}, and (ii) couple to angular momentum operators respecting multistate resolutions by definition~\cite{Xiao2010}. As such, the multistate topologies should be in principle anomalously manifested, and hence observable, through certain magnetic responses.

\begin{figure}[t!]
    \centering
    \includegraphics[width=0.9\linewidth]{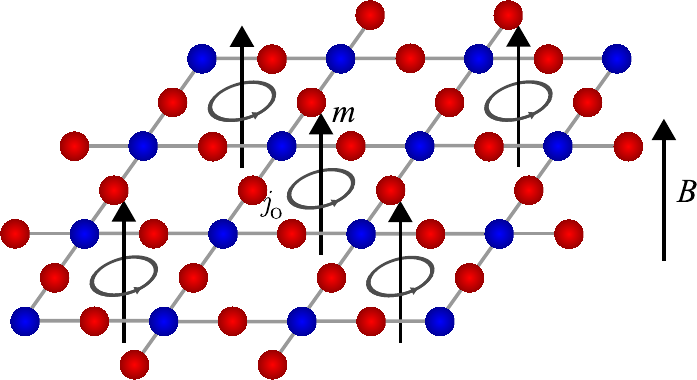}
    \caption{Orbital magnetization $(\textit{m})$ from loop currents $(j_{\text{O}})$, in~response to an out-of-plane magnetic field $(\textit{B})$. The orbital response fingerprints the nontrivial Euler band topology.}
    \label{fig1}
\end{figure}

In this work, we identify the role of multistate topological invariants in orbital magnetic susceptibilities and responses. We find a rich interplay. First, we show that the orbital magnetic susceptibility can be naturally decomposed in terms of energetic and quantum geometric contributions \cite{Gao2015, Ogata2015}. The former can be estimated directly from the energy spectrum and fermionic occupation numbers determined by the quasiparticle energies and chemical potential. Thus, the energetic contribution can be deduced solely from the angle-resolved photoemission spectroscopy (ARPES) experiments. On the other hand, we find that the geometric term explicitly involves the quantum metric tensor and encodes the information about the topological invariants realized by electronic states in the crystalline material. Second, we leverage this interplay of energetic and geometric orbital magnetic susceptibility terms to reconstruct multiband topological invariants in continuum and lattice models with multiband Euler topology. In particular, we detail this correspondence for Euler topologies that realize Euler band nodes. Fundamentally, the \mbox{Euler} nodes act as sources of non-Abelian Berry flux associated with multiband Berry connections, which is distinct from standard Abelian Berry flux monopoles supported by intraband Berry connections. Beyond the orbital magnetic susceptibility features induced by the multiband Euler invariants, as identified in this work, the Euler class can be in principle alternatively deduced from optical absorptivities~\cite{Jankowski2025PRBoptical} and nonlinear anomalous Hall transport phenomena~\cite{Jain2025} in responses to electric, rather than magnetic fields. Third, we showcase the correspondence of orbital magnetic susceptibility and multiband topological invariant in a material realization context of strontium ruthenate ($\text{Sr}_2 \text{Ru} \text{O}_4$) models. Therefore, our findings not only demonstrate that quantum geometry plays an important role in orbital magnetic susceptibility, but also allows one to experimentally reveal the presence of exotic multiband topological invariants realized in quasiparticle bands. Hence, we establish the orbital magnetic susceptibility as an observable of interest for probing multiband topologies, analogously to the anomalous Hall conductance constituting the hallmark observable for single-band Chern invariants in topological \mbox{matter}. \mbox{Notably}, unlike~the bulk charge conductivity due to the single-band \mbox{invariants}, which can be related to the edge conductivities in Chern \mbox{insulators}, the identified orbital magnetic susceptibility due to the multiband invariants is not reflected by any \mbox{universal} edge effects.

In the following, we subsequently detail how the orbital magnetic susceptibility (Sec.~\ref{sec::II}) arises as a probe of multiband topology (Sec.~\ref{sec::III}), and captures the multiband quantum geometry even in effective models of real materials. We further discuss these findings (Sec.~\ref{sec::IV}), before concluding (Sec.~\ref{sec::V}).

\section{Orbital magnetic susceptibility}\label{sec::II} We now demonstrate how the multiband topology can be revealed via orbital magnetization measurements in responses of quasi-two-dimensional crystals. Our findings lie beyond the limits of previously studied magnetic properties and responses of Landau levels~\cite{Rhim2020}, see Appendix~\ref{app::A} for comparison. The orbital magnetization $m$ is given by an orbital susceptibility $\chi_\text{O}$, $m = \chi_\text{O} B$, upon applying an external out-of-plane magnetic field $B$ (see Fig.~\ref{fig1}). In terms of current velocity \mbox{operators} $j_x,\;j_y$, the orbital magnetic susceptibility $\chi_\text{O}$ is given by Fukuyama formula \cite{Fukuyama1971} with correction terms \cite{Raoux2015, Gomez2011}:
\beq{}
    \chi_\text{O} = \frac{e^2}{2\hbar^2 c^2 }k_BT\sum_{n,\kv} \Tr[G j_x G j_y G
    (j_x G j_y+ j_{xy})] + (x\leftrightarrow y)
\eec
where we sum over the Matsubara frequencies $\omega_n$ and single-particle momenta $\textbf{k}$. ${j_{xy} =\partial _{k_x}\partial_{k_y} H}$, and ${G = 1/(i \omega_n + \mu - H)}$ is an electronic Green’s function $G$, given by Hamiltonian $H$ and chemical potential $\mu$. Centrally to this work, we recast the orbital magnetization in terms of gauge-independent energetic ($\chi_\text{E}$) and geometric contributions ($\chi_\text{geo}$), $\chi_\text{O} = \chi_\text{E} + \chi_\text{geo}$.
The energetic orbital susceptibility term $\chi_\text{E}$ in the decomposition can be identified as the Landau-Peierls formula~\cite{Peierls1933},
\beq{eq:chiE}
    \chi_\text{E} 
    =
    \frac{e^2}{6 \hbar^2 c^2} \sum_a\sum_\kv
    \left[
    \frac{\partial^2\ep_a}
    {\partial k_x^2}
    \frac{\partial^2\ep_a}
    {\partial k_y^2}
    -
    \lb
    \frac{\partial^2\ep_a}
    {\partial k_x\partial k_y}
    \rb^2
    \right]
    f'_a
\eec
which can be directly estimated from the energy spectrum ($\ep_a$) measurable in ARPES experiments. $a$ labels the quasiparticle bands, and $f_a(\mu) = \big[\text{exp}(\frac{\ep_a - \mu}{k_B T})+1\big]^{-1}$ is the Fermi-Dirac distribution function, which establishes equilibrium occupations. The \mbox{quantum} geometric contribution to orbital susceptibility~\cite{Piechon2016}, $\chi_{\mathrm{geo}}$, can be decomposed in terms of three geometric terms, ${\chi_{\mathrm{geo}} = \chi_{xx}+\chi_{yy}+\chi_{xy}}$, which read 
\begin{align}
    \chi_{xx}
    &=
    \frac{e^2}{\hbar^2 c^2}\sum_{a\neq b}\sum_\kv
    \partial_y\ep_a
    (\partial_y\ep_a + \partial_y\ep_b) g^{ab}_{xx} 
    \lb
    f'_a-\frac{f_{ab}}{\ep_{ab}}
    \rb\;,\label{eq:chixx}\\
    \chi_{yy}
    &=
    \frac{e^2}{\hbar^2 c^2}\sum_{a\neq b}\sum_\kv
    \partial_x\ep_a
    (\partial_y\ep_a + \partial_x\ep_b) g^{ab}_{yy} 
    \lb
    f'_a-\frac{f_{ab}}{\ep_{ab}}
    \rb\;,\label{eq:chiyy}
\end{align}
\begin{align}
    \chi_{xy}
    &=
    \frac{e^2}{\hbar^2 c^2}\sum_{a\neq b}\sum_\kv 
    g^{ab}_{xy}\left[
    \ep_{ab}
    \partial_x\ep_a\partial_y\ep_af_{a}''
    -
    \lb
    f'_a-\frac{f_{ab}}{\ep_{ab}}
    \rb
    \right.
    \nonumber\\
    &\left.
    \phantom{\frac{1}{2}}
    (2\partial_x\ep_a\partial_y\ep_a
    +
    \partial_x\ep_a\partial_y\ep_b
    +
    \partial_y\ep_a\partial_x\ep_b
    -
    \ep_{ab}\partial_x\partial_y\ep_a)
    \right]
    \;.\label{eq:chixy}
\end{align}
Here, $g^{ab}_{\mu \nu} = \Re[\braket{\partial_\mu u_{a\kv}}{u_{b\kv}} \braket{u_{b\kv}}{\partial_\nu u_{a\kv}}]$ is a multiband quantum metric tensor~\cite{Ahn2020, Ahn2021, Bouhon2023, Jankowski2025PRBoptical, Jain2025} for Bloch quasiparticle states, $\ket{\psi_{a\kv}} = e^{i \kv \cdot \textbf{r}} \ket{u_{a\kv}}$, $\partial_\nu \equiv \frac{\partial}{\partial{k_\nu}}$, ${\epsilon_{ab} \equiv \epsilon_a - \epsilon_b}$, and ${f_{ab} \equiv f_a - f_b}$. For a detailed derivation of the energetic and quantum geometric contributions to the orbital magnetic susceptibility $\chi_\text{O}$, including the higher-order corrections, see the Appendix~\ref{app::B}.

\begin{figure}[t]
    \centering
    \includegraphics[width=\linewidth]{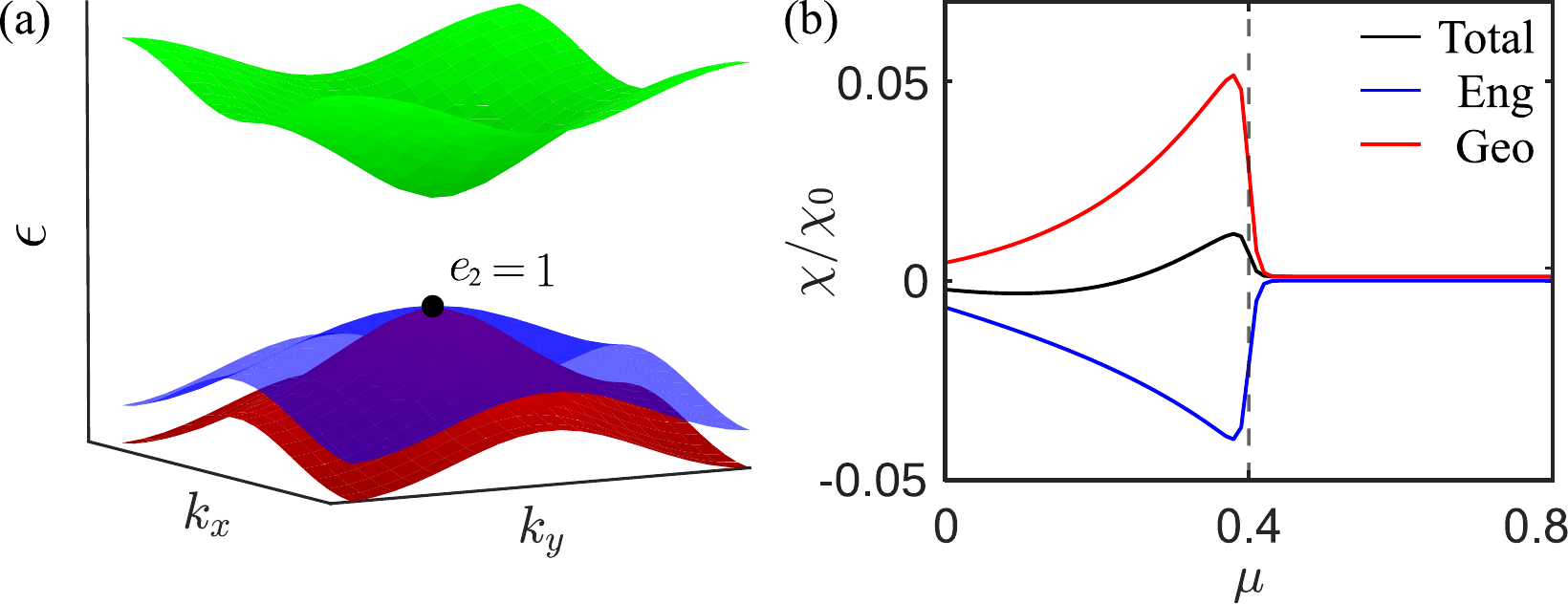}
    \caption{Topological bands with Euler charge $e_2=1$ in Lieb lattice model. {\bf (a)} Low-energy bands supporting the Euler invariant via a quadratic band touching. {\bf (b)} The quantum geometric contribution to orbital magnetic susceptibility associated with the Euler charge ($\text{Geo)}$ is of opposite sign to the band-dispersive susceptibility contribution ($\text{Eng})$. The dashed line indicates the energy of the band touching with nontrivial Euler class.}
    \label{fig2}
\end{figure}

\section{Multiband topologies in orbital magnetization}\label{sec::III} 

We now showcase how the geometric orbital magnetic susceptibility term $\chi_\text{geo}$ allows to deduce the Euler class invariant $e_2 \in \mathbb{Z}$~\cite{Bouhon2019, Ahn2019}. The Euler invariant $e_2$ is defined using the multiband connection between bands $a$ and $b$ expressed in a real gauge, ${A}_\mu^{ab} = \langle u_{a\kv} \vert \partial_\mu u_{b \kv} \rangle$, which determines the Euler form ${\mathrm{Eu}^{ab} = \partial_x {A}_y^{ab} - \partial_y {A}_x^{ab}}$~\cite{Ahn2019, Bouhon2019nonabelian}. The multiband Euler invariant over a two-dimensional Brillouin zone (BZ) patch $\Df$ reads~\cite{Ahn2019},
\beq{Eu_Patch}
    e_2 = 
    \frac{1}{2\pi}\left[
        \int_\Df \text{d}^2 \kv~\mathrm{Eu}^{ab}
        - \oint_{\partial\Df} \text{d}\kv \cdot {\textbf{A}}^{ab}
    \right]\; \in \mathbb{Z},
\eeq
and captures an obstruction to Stokes's theorem applied to the multiband connection. The topology of an Euler node, as an obstruction to Stokes' theorem, can be furthermore formulated in terms of path-ordered Wilson loops~\cite{Bouhon2019, Bouhon2019nonabelian}. The quantization of the invariant is protected by a spatiotemporal inversion ($\mathcal{PT}$) symmetry~\cite{Bouhon2019, Ahn2019, Bouhon2019nonabelian}. The minimal integer Euler invariant, $|e_2|=1$, can be realized by a quadratic band touching within a general class of $\kv\cdot\mathbf{p}$ models expressed in terms of band projector matrices $P_1, P_2$,
\begin{align}
    H_{\text{Eu}} &= \frac{k^2}{2m_1}P_1 + \frac{k^2}{2m_2}P_2
   = \frac{1}{4m_1m_2}[(m_1+m_2)k^2\s_0 \nonumber\\&\:+2(m_1-m_2)k_xk_y\s_1+(m_1-m_2)(k_x^2-k_y^2)\s_3]\ .\label{HP_Eu}
\end{align}
$m_1$ and $m_2$ are the effective masses of the touching bands, $k^2 = k_x^2 +k^2_y$, and $\s_i$ are the Pauli matrices. In Fig.~\ref{fig2}(a), we present a~quadratic band touching realizing an Euler class charge $|e_2|=1$ in bands with effective masses $m_1,m_2 < 0$. In Fig.~\ref{fig2}(b), we demonstrate the corresponding orbital magnetic susceptibility $\chi_\text{O}$, decomposed in terms of $\chi_\text{E}$ and $\chi_\text{geo}$ contributed by the multiband quantum metric $g_{\mu \nu}^{ab} = \text{Re}({A}_\mu^{ab}  {A}_\nu^{ba})$. The connection between the geometric part of the orbital susceptibility~$\chi_\text{geo}$ and a violation of the Stokes' theorem by the Euler node lies in the associated singular behavior of the multiband quantum metric $g^{ab}_{\mu\nu}$, which enters the orbital magnetic susceptibility [\mbox{Eqs.~(\ref{eq:chixx}-\ref{eq:chixy})]}. For full analytical derivation, see~Appendixes~\ref{app::B},~\ref{app::C}. 

We note that the topologically induced geometric contribution $(\chi_\text{geo})$ opposes the energetic contribution $(\chi_\text{E})$, leading to a potential sign reversal in the total orbital magnetic susceptibility $(\chi_\text{O})$, depending on the relative strength of each contributions.

Furthermore, beyond the general effective $\kv\cdot\mathbf{p}$ expansions, we demonstrate an emergence of analogous multiband topological quantum geometric susceptibility contributions in lattice-regularized Lieb lattice models and tight-binding Hamiltonian of strontium ruthenate ($\text{Sr}_2 \text{Ru} \text{O}_4$)~\cite{Noce1999}, fitted to the theoretical band energies of the experimentally accessible quasiparticle energies measured in ARPES. 

\subsection{Lieb lattice models}

We here detail the Lieb lattice models with multiband Euler topology ($e_2 \neq 0$).
The corresponding Lieb lattice Hamiltonians are constituted by $s$-orbitals at all lattice sites, involve nearest-neighbor hoppings ($t$) and onsite energies ($M$) \cite{Chau2025}. In momentum~space basis, the considered Bloch Hamiltonians $h_{\text{Lieb}}(\kv)$ read:
\begin{align}
    h_{\text{Lieb}}(\kv) &= 
    \begin{pmatrix}
        M   &   2t\cos{(k_x/2)} &   2t\cos{(k_y/2)} \\
        2t\cos{(k_x/2)} &0  &0  \\
        2t\cos{(k_y/2)} &0  &0  
    \end{pmatrix}\ .
\end{align}
For $M\neq 0$, $h_{\text{Lieb}}(\kv)$ realizes an Euler node at the $K$ point, with patch Euler class of $|e_2|=1$, calculated using Eq.~\eqref{Eu_Patch}. In Fig.~\ref{fig2}, we show the orbital magnetic susceptibility with geometric contributions due to the nontrivial multiband Euler topology.

\subsection{Sr$_2$RuO$_4$}

For Sr$_2$RuO$_4$, one can treat the system as pseudo-two-dimensional, and decouple the Hamiltonian into two parts: in-plane, and out-of-plane \cite{Noce1999}. The in-plane part is constituted by the Lieb lattice with orbitals ($\mathrm{Ru}\;d_{xy},\;\mathrm{O}1\;p_{x},\;\mathrm{O}2\;p_{y}$):
\beq{Sr2RuO4_xy}
    h_{\mathrm{Sr_2RuO_4},xy}
    =
    \begin{pmatrix}
        \ep_{xy} & it_1(k_x) & it_1(k_y)
        \\
        -it_1(k_x) & \ep_p & t_2(k_x,\;k_y)
        \\
        -it_1(k_y) & t_2(k_x,\;k_y) & \ep_p
    \end{pmatrix}
\eec
where the off diagonal elements are $t_1(k_\mu)=-2t_1\sin (k_\mu)$ and $t_2(k_x,\;k_y)=-4t_2\sin(k_x/2)\sin(k_y/2)$. We note that only the isolated band is relevant at the Fermi surface $\mu=0$, thus the magnetic signature of the Euler node cannot be directly observed from the in-plane part. The out-of-plane part consists of four orbitals ($\mathrm{Ru}\;d_{xz},\;\mathrm{Ru}\;d_{yz},\;\mathrm{O}1\;p_{z},\;\mathrm{O}2\;p_{z}$):
\beq{Sr2RuO4_z}
    h_{\mathrm{Sr_2RuO_4},z}
    =
    \begin{pmatrix}
        \ep_{d} & 0 & it_3(k_x) & 0
        \\
        0 & \ep_d & 0 & it_3(k_y)
        \\
        -it_3(k_x) & 0 & \ep_p & t_4(k_x,\;k_y)
        \\
        0 & -it_3(k_y) & t_4(k_x,\;k_y) & \ep_p
    \end{pmatrix}
\eec
where the off-diagonal elements are $t_3(k_\mu)=-2t_3\sin (k_\mu)$ and $t_4(k_x,\;k_y)=-4t_4\cos(k_x/2)\cos(k_y/2)$.
Due to the presence of $\mathcal{PT}$ symmetry, the Hamiltonian can be converted to the real gauge with a unitary:
\beq{}
    U = \begin{pmatrix}
        1 & 0 & 0 & 0 \\
        0 & 1 & 0 & 0\\
        0 & 0 & 0 & -i \\
        0 & 0 & i & 0\\
    \end{pmatrix},
\eeq
which allows us to compute the patch Euler class, $e_2 \in \mathbb{Z}$, using Eq.~\eqref{Eu_Patch}. We note that $h_{\mathrm{Sr_2RuO_4},z}$ hosts two separated pairs of bands. For the pair of bands near the Fermi surface, it host two Euler nodes, each of patch Euler class $|e_2|=1$, respectively at the $\G$ and $K$ points. The tight-binding parameters for the Hamiltonian are obtained from Ref.~\cite{Noce1999}, and are summarized as a table below in the units of eV:
\\
\begin{small}
\begin{equation*}
    \begin{tabular}{|c|c|c|c|c|c|c|c|}
        \hline\
        Tight-binding parameters &
        $\ep_{xy}$ & $\ep_d$ & $\ep_p$ & $t_1$ & $t_2$ & $t_3$ & $t_4$\\
        \hline
        Values (eV) &
        --1.9 & --0.9 & --2.4 & 1.1 & --0.52 & 0.85 & 0.1
        \\
        \hline
    \end{tabular}.
\end{equation*}
\end{small}
\\

\begin{figure}[t!]
    \centering
    \includegraphics[width=\linewidth]{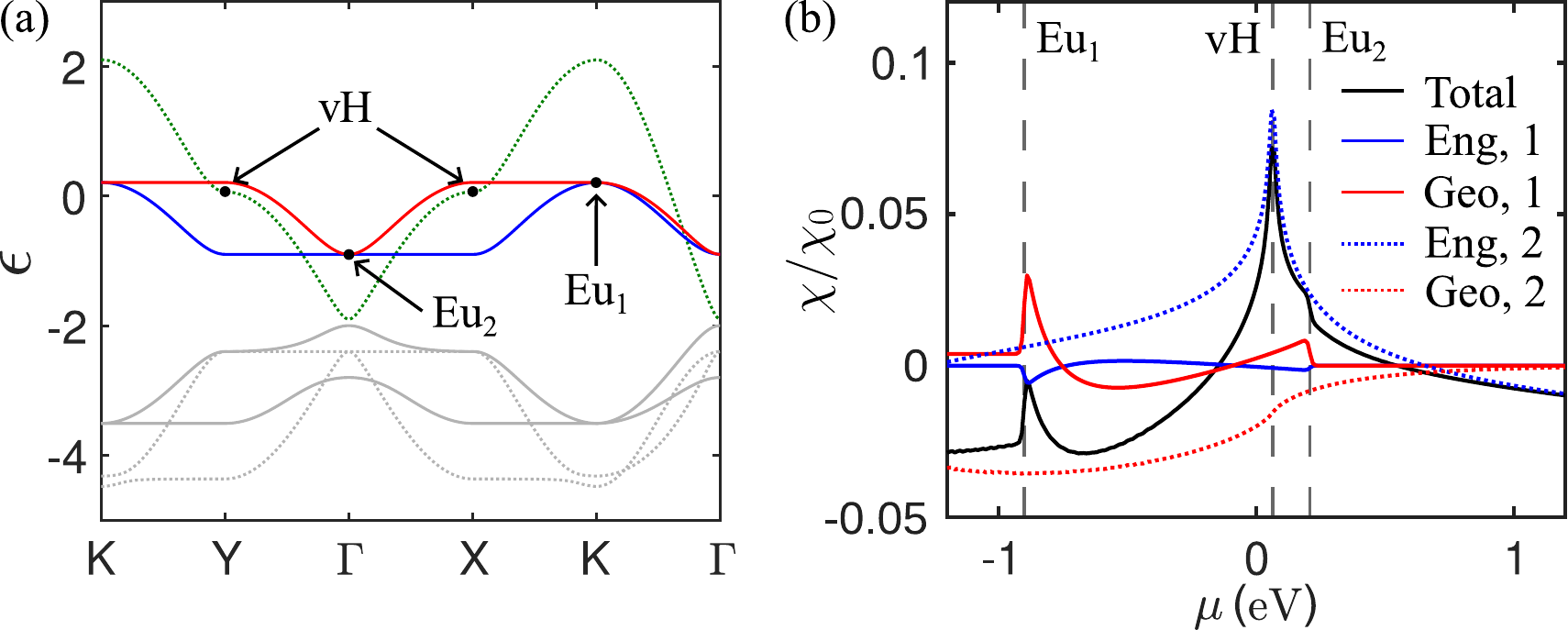}
    \caption{Orbital magnetic susceptibility fingerprints topological Euler invariants in Sr$_2$RuO$_4$ model.
    {\bf (a)} Band structure obtained from a tight-binding model of Sr$_2$RuO$_4$. Bands contributed by the $(\mathrm{Ru}\;d_{xz},\;\mathrm{Ru}\;d_{yz},\;\mathrm{O}1\;p_{z},\;\mathrm{O}2\;p_{z})$ orbitals (solid lines) realize the nontrivial Euler class invariants, $|e_2| =1$ $(\text{Eu}_{1,2})$; dotted lines mark $(\mathrm{Ru}\;d_{xy},\;\mathrm{O}1\;p_{x},\;\mathrm{O}2\;p_{y})$ orbital bands hosting van Hove singularities ($\text{vH}$).
    {\bf (b)} Orbital magnetic susceptibility $\chi$ decomposed into energetic (Eng) and geometric (Geo) contributions as a function of chemical potential $\mu$, with $\chi_0= t_0(e a_{0}^2/4\pi^2\hbar c)^2$ where $a_{0}$ is the lattice constant and $t_0 = 1$\;eV. Around $\mu = -0.8$ and $\mu = 0.2~\text{eV}$, the total orbital magnetic susceptibility $\chi_\text{O}$ reflects the presence of topologically induced concentrated quantum geometries due to the Euler charges.}
    \label{fig3}
\end{figure}

In Fig.~\ref{fig3}(a), we show the electronic bands within the~seven-band model of Sr$_2$RuO$_4$, where we retrieve the Euler invariants $|e_2|=1$ in the higher energy band subspace close to the Fermi level. Figure~\ref{fig3}(b) demonstrates the geometric orbital susceptibility contributions contributed by the Euler charges in distinct band subspaces. Once again, we find that $\chi_\text{O}$ depends heavily on the topologically induced singular contribution of quantum metric $g_{\mu \nu}^{ab}$, which is realized between distinct pairs of bands contributing to $\chi_\text{geo}$. The remaining enhanced contribution to orbital susceptibility in Fig.~\ref{fig3}(b) arises from van Hove (vH) singularities, rather than being a multiband topological effect. This results in a combination of enhanced density of states and the corresponding occupation factors. However, the vH singularity contributions can be disregarded as nontopological purely from the inspection of ARPES spectra. Thus, upon disregarding the expected vH contributions, $\chi_\text{O}$~allows reconstructing the nontrivial multistate geometry and multiband topologies purely from $\chi_\text{geo}$ and its dependence on doping ($\mu$), as we address next.

\section{Discussion}\label{sec::IV}  
\begin{figure}[t!]
    \centering
    \includegraphics[width=\linewidth]{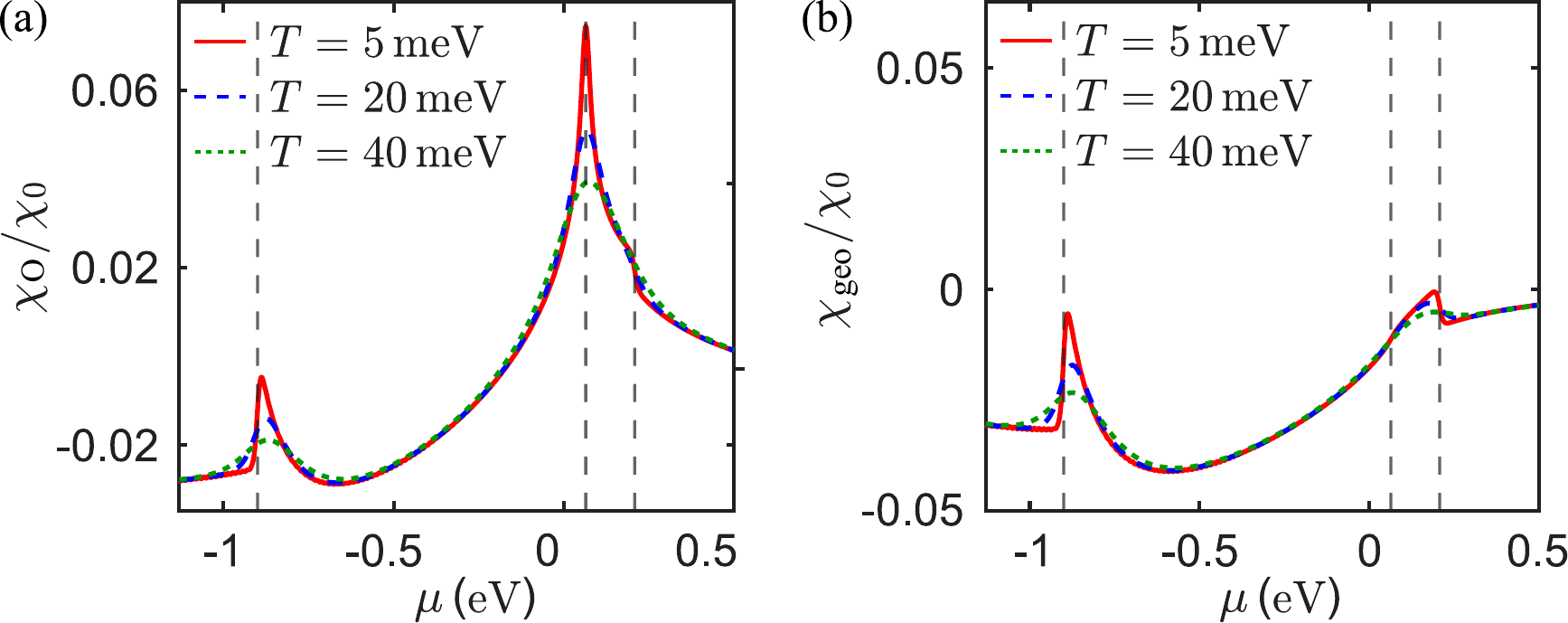}
    \caption{Thermal dependence of the orbital magnetic susceptibility due to an Euler invariant, $|e_2| = 1$. {\bf(a)} Total orbital magnetic susceptibility $\chi_\text{O}$, {\bf(b)} geometric contribution $\chi_{\text{geo}}$ at increasing finite temperatures $T = 5,\;20,\; 40$ meV in $\mathrm{Sr_2RuO_4}$ model, as a function of chemical potential $\mu$, with ${\chi_0= t_0(e a_0^2/4\pi^2\hslash c)^2}$, where $a_0$ is the lattice constant and $t_0 = 1$\;eV.}
    \label{fig4}
\end{figure}

We further discuss how the multiband topological invariant can be deduced from the orbital magnetization response. Furthermore, we show the thermal dependence of the geometric orbital magnetic susceptibility in Fig.~\ref{fig4}, which demonstrates the robustness of the orbital magnetization features as indicators of the multiband Euler topology. Moreover, we provide a~semiclassical interpretation of our findings, which we phrase in terms of the multiband wave~packet dynamics.

To reconstruct the multiband invariant, one may first identify possible quantum geometry hotspots, such as band touchings observable in ARPES. In the case of trivial multiband invariants, only the energetic orbital susceptibility contribution $\chi_\text{E}$ arises. $\chi_\text{E}$ can be directly estimated based on the ARPES band structure data, from Eq.~\eqref{eq:chiE}. Importantly, the ARPES data on its own is insufficient to deduce a presence of an Euler node, as any quadratic band touching possibly measured via photoemission cannot be distinguished from being realized via an effective Hamiltonian: $H_\text{eff}(\textbf{k}) =\frac{k^2}{2m} \sigma_3$. Such Hamiltonian realizes a quadratic dispersion, yet, corresponds to a trivial Euler invariant $e_2 = 0$. We propose that the geometric contribution $\chi_\text{geo}$ can be thus extracted from combining ARPES and orbital magnetic susceptibility $\chi_\text{O}$ data with the $\chi_\text{E}$ estimation. Importantly, the known spin magnetic susceptibility contributions must be further accounted for, when the total magnetic susceptibility is measured~\cite{Oh2025}. Notably, the spin magnetic susceptibility contribution can be measured directly via Knight shift using nuclear magnetic resonance~\cite{Ishida1998}. As such, the spin contribution can be traced out from the total susceptibility, which allows one to access the orbital contribution studied in this work.

To obtain the geometric part of the orbital susceptibility ($\chi_\text{geo}$)  experimentally, (i) the total magnetic susceptibility should be measured, (ii) the spin contribution must be excluded, and (iii) the energetic contribution must be finally subtracted. The presence of a~multiband topological invariant in experiment can be validated upon inserting an analytical ansatz for $g^{ab}_{\mu \nu}~\propto |e_2|^2/|k|^2$ in Eqs.~(\ref{eq:chixx}-\ref{eq:chixy}), which follows from Eq.~\eqref{HP_Eu}, in the momentum space neighborhood of the band touchings. The contribution arising from this ansatz can be compared against the experimental data.

It should be noted that although a resolution of the quantum metric tensor in momenta has been experimentally achieved in solid-state materials very recently~\cite{Kang2025, Kim2025}, these findings are limited by a two-band approximation. In this work, we go beyond this paradigm. Namely, we demonstrate that the multiband metric and topological invariants can be inferred even in a seven-band model of $\text{Sr}_2 \text{Ru} \text{O}_4$, assuming the full knowledge of the ARPES spectrum, as in the mentioned experiments. 

Finally, we provide a semiclassical interpretation of our findings. The orbital susceptibility contributed by quantum geometry reflects orbital currents, which semiclassically arise from the loop current dynamics of wave packets. Increasing magnetic field $B$ induces a Lorentz force on the wave~packet, which responds accordingly. The response arises from a field-induced positional shift correction in Berry connection, $A'_i(B) = F_{ij} B_j$, with $F_{ij}$, the anomalous orbital polarizability tensor introduced by Gao and Niu \cite{Gao2014, Gao2015}. The $F_{ij}$ tensor involves the multiband connections, which connect the response directly to the Euler curvature form, and can support orbital magnetization via field-induced Berry curvature, $\Om'_{xy}(B) = \partial_x A'_y(B) - \partial_y A'_x(B)$. The associated positional wave~packet shift results in orbital currents, as follows from a direct inspection. We leave the further interplay of magnetic field-induced positional shifts and multiband geometries for future work~\cite{Chau2026MNLHE}.

\section{Conclusion}\label{sec::V} To sum up, we demonstrate that the presence of multiband Euler topology in a bulk material can be inferred from its orbital magnetization responses. Our findings establish orbital magnetization as an experimentally measurable physical property of choice for identifying the quantum geometric fingerprint of multiband topological Euler invariant. We thus propose an electromagnetic smoking-gun probe for pairs of topological bands supporting vorticity in a multiband Berry connection. Our work establishes a physical go-to response for probing Euler bands, analogously to the transverse Hall conductance as the hallmark response of Chern bands realizing intraband Berry curvature in quantum Hall phases.

\section*{Acknowledgments} The authors thank Fr\'ed\'eric Pi\'echon and Gaurav Chaudhary for discussions and helpful comments on the manuscript. C.W.C. \mbox{acknowledges} funding from the Croucher Cambridge International Scholarship by the Croucher Foundation and the Cambridge Trust. \mbox{R.-J.S.} acknowledges funding from an EPSRC ERC underwrite Grant No.~EP/X025829/1, and a Royal Society exchange Grant No. IES/R1/221060, as well as Trinity College, Cambridge. W.J.J. acknowledges funding from the Rod Smallwood Studentship at Trinity College, Cambridge. This research was supported in~part by grant NSF PHY-2309135 to the Kavli Institute for Theoretical Physics (KITP).

\appendix

\begin{widetext}

\section{Magnetic response function of Euler bands}\label{app::A}
In this section, we consider the implication of quantum geometry on the magnetic response function of Euler bands. Using the Roth-Gao-Niu relation~\cite{Gao2017, Fuchs2018, Roth1966} for electrons in magnetic field $B$ at zero temperature ($T = 0$):
\beq{RGN}
    \lb n+\frac{1}{2}\rb\frac{eB}{h}=N_0(\mu)+B\partial_\ep M_0|_{\ep=\mu}+\frac{1}{2}B^2\partial_\ep\chi_0|_{\ep=\mu}+\bigO(B^3)
\eec
where we consider a half-filled Landau level, such that the Fermi energy $\mu$ matches the $n$\text{th} Landau level energy. $N_0(\mu)$ is the integrated density of states at zero magnetic field, $M_0(\mu)$ is the spontaneous magnetization, and $\chi_0(\mu)$ is the magnetic susceptibility. 
\subsection{Landau level}
We begin by discussing the Landau level for a continuum Hamiltonian of an Euler node. Around an Euler node with $|e_2|=1$, one can write down an effective two-band $\kv\cdot\mathbf{p}$ model using projector matrices:
\begin{align}
    P_1 &= \begin{pmatrix}
        \sin^2\th & -\sin\th\cos\th
        \\
        -\sin\th\cos\th & \cos^2\th
    \end{pmatrix}\ ,
    \\
    P_2 &= \begin{pmatrix}
        \cos^2\th & \sin\th\cos\th
        \\
        \sin\th\cos\th & \sin^2\th
    \end{pmatrix}\ ,
    \\
    H_{\text{Eu}} &= \frac{k^2}{2m_1}P_1 + \frac{k^2}{2m_2}P_2
    \nonumber\\
    &= \frac{1}{4m_1m_2}[(m_1+m_2)k^2\s_0
    +2(m_1-m_2)k_xk_y\s_1+(m_1-m_2)(k_x^2-k_y^2)\s_3].
\end{align}
$m_1$ and $m_2$ are the effective masses of the touching quadratic bands, $k^2 = k_x^2 +k^2_y$, $\th =\arg(k_x + ik_y)$ and $\s_i$ are the Pauli matrices.
We now perform a replacement $k_x\ra (a+a^\dagger)/\sqrt{2}l_B$ and $k_y\ra i(a-a^\dagger)/\sqrt{2}l_B$ \cite{Rhim2020}, with the magnetic length $l_B=\sqrt{\hbar/eB}$ and $a~(a^\dagger)$ being the annihilation~(creation) operator. We further ensure that the Hamiltonian is Hermitian by symmetrizing $k_x k_y$ as:
\begin{align}
    k_xk_y &= \frac{1}{2}(k_xk_y+k_yk_x)\nonumber\\
    &=\frac{i}{2l_B^2}(aa-a^\dagger a^\dagger)\ ,
    \\
    k^2 &= \frac{1}{l_B^2}(aa^\dagger + a^\dagger a)\ ,
    \\
    k_x^2-k_y^2 &= \frac{1}{l_B^2}(aa + a^\dagger a^\dagger)\ ,
\end{align}
where for reference we have included the form of $k^2$ and $k_x^2-k_y^2$. We thus obtain a Landau level Hamiltonian ($H_{\text{Eu},\text{LL}}$):
\begin{align}
    H_{\text{Eu}}&=\frac{1}{4m_1m_2}[(m_1+m_2)k^2\s_0+2(m_1-m_2)k_xk_y\s_1+(m_1-m_2)(k_x^2-k_y^2)\s_3]\nonumber\\
    H_{\text{Eu},\text{LL}}&=\frac{1}{4m_1m_2}\left[(m_1+m_2)\frac{1}{l_B^2}(aa^\dagger + a^\dagger a)\s_0
    +(m_1-m_2)\frac{i}{l_B^2}(aa-a^\dagger a^\dagger)\s_1
    +(m_1-m_2)\frac{1}{l_B^2}(aa + a^\dagger a^\dagger)\s_3\right]\nonumber\\ 
    &=\frac{1}{4m_1m_2l_B^2}\left[(m_1+m_2)(aa^\dagger + a^\dagger a)\s_0
    +i(m_1-m_2)(aa-a^\dagger a^\dagger)\s_1
    +(m_1-m_2)(aa + a^\dagger a^\dagger)\s_3\right]\ ,\label{hLL_Eu}
\end{align}
which can be solved using a wavefunction ansatz:
\beq{}
    \ket{\psi}=\sum^\infty_{n=0}v_n\ket{n}.
\eeq

Here, $\ket{n}$ are normalized states that satisfy $a^\dagger\ket{n} = \sqrt{n+1}\ket{{n+1}}$ and $a\ket{n} = \sqrt{n}\ket{{n-1}}$. As such, we have the following relations:
\begin{align}
    (aa^\dagger + a^\dagger a)\ket{n} 
    &=
    a\sqrt{n+1}\ket{{n+1}} + a^\dagger\sqrt{n}\ket{{n-1}}
    \nonumber\\
    &=
    (2n+1)\ket{n}\ ,
    \\
    (aa + a^\dagger a^\dagger)\ket{n} 
    &=
    a\sqrt{n}\ket{{n-1}} + a^\dagger\sqrt{n+1}\ket{{n+1}}
    \nonumber\\
    &=
    \sqrt{n(n-1)}\ket{n-2} + \sqrt{(n+1)(n+2)}\ket{n+2}\ ,
    \\
    (aa - a^\dagger a^\dagger)\ket{n}
    &=
    \sqrt{n(n-1)}\ket{n-2} - \sqrt{(n+1)(n+2)}\ket{n+2}\ ,
\end{align}
which gives a Hamiltonian:
\beq{HLL_Eu}
    H_{\text{Eu},\text{LL}} = 
    \begin{pmatrix}
        h_0& 0 & g_0 & 0 & \cdots\\
        0& h_1 & 0 & g_1 & \cdots\\
        g_0^\dagger& 0 & h_2 & 0 & \cdots\\
        0 & g_1^\dagger & 0 & h_3 & \cdots\\
        \vdots & \vdots &\vdots &\vdots & \ddots
    \end{pmatrix}
\eec
where:
\begin{align}
    h_n &= \a(2n+1)(m_1+m_2)\s_0\ ,
    \\
    g_n &= \a\sqrt{(n+1)(n+2)}(m_1-m_2)(\s_3-i\s_1)\ .
\end{align}
Above, we have defined $\a=(4m_1m_2l_B^2)^{-1}$. To determine $v_n$ thus the Landau level wavefunction, one first notes the following:
\begin{align}
    (\s_3\pm i\s_1)v_\pm
    &=
    2v_\mp\ ,
    \\
    (\s_3\pm i\s_1)v_\mp
    &=
    0\ ,
    \\
    v_\pm &= \frac{1}{\sqrt{2}}\begin{pmatrix}
        i \\ 1
    \end{pmatrix}\ .
\end{align}
This allows us to write down an exact solution for the wavefunction for $\ket{\psi_n}$, which is a linear combination of $\ket{n}$ and $\ket{n-2}$ for $n\geq 2$, as $g_nv_+ =0$ and $g_n^\dagger v_-=0$. Specifically, we have:
\begin{align}
    \ket{\psi_n} &= c_n v_+ \ket{n} + d_n v_- \ket{n+2}\; ,
    \\
    H_{\text{Eu},\text{LL}} \ket{\psi_n} &= E_{\text{Eu},\text{LL},n}\ket{\psi_n}
    \nonumber\\
    \text{LHS}&=
    \begin{pmatrix}
        h_{n-4} & g_{n-4} & 0 & 0 \\
        g^\dagger_{n-4} & h_{n-2} & g_{n-2} & 0 \\
        0 & g_{n-2}^\dagger & h_n & g_n \\
        0 & 0 & g_n^\dagger & h_{n+2}
    \end{pmatrix}
    \begin{pmatrix}
        0 \\ c_nv_+ \\ d_nv_- \\ 0
    \end{pmatrix}
    \nonumber\\
    &=
    \begin{pmatrix}
        c_ng_{n-4}v_+
        \\
        c_nh_{n-2}v_+ + d_ng_{n-2}v_-
        \\
        c_ng_{n-2}^\dagger v_+ + d_nh_nv_-
        \\
        d_n g_n^\dagger v_-
    \end{pmatrix}
    \nonumber\\
    &=
    \a
    \begin{pmatrix}
        0
        \\
        [c_n(2n-3)(m_1+m_2) + 2d_n\sqrt{n(n-1)}(m_1-m_2)]v_+
        \\
        [2c_n\sqrt{n(n-1)}(m_1-m_2) + d_n(2n+1)(m_1+m_2)]v_-
        \\
        0
    \end{pmatrix}
    \nonumber\\
    E_{\text{Eu},\text{LL},n}\begin{pmatrix}
        c_n \\ d_n
    \end{pmatrix}
    &=
    \a \begin{pmatrix}
        (2n-3)(m_1+m_2) & 2(m_1-m_2)\sqrt{n(n-1)} \\
        2(m_1-m_2)\sqrt{n(n-1)} & (2n+1)(m_1+m_2)
    \end{pmatrix}
    \begin{pmatrix}
        c_n \\ d_n
    \end{pmatrix}
    \; ,
    \\
    E_{\text{Eu},\text{LL},n,\pm} &= \a\left[(m_1+m_2)(2n-1)\pm 2\sqrt{(m_1+m_2)^2+n(n-1)(m_1-m_2)^2}\right]
    \nonumber\\
    &=
    2\a(m_1+m_2)\left[n-\frac{1}{2}\pm\sqrt{1+n(n-1)\lb\frac{m_1-m_2}{m_1+m_2}\rb^2}\right]\label{E_n}\;.
\end{align}
For $n=0$ and $n=1$, these need to be solved separately, where $\ket{\psi_0}\propto v_-\ket{0}$ and $\ket{\psi_1}\propto v_-\ket{1}$, giving:
\begin{align}
    E_{\text{Eu},\text{LL},0} &= \a(m_1+m_2) \;,\\
    E_{\text{Eu},\text{LL},1} &= 3\a(m_1+m_2) \;.
\end{align}
For trivial case, $g_n=0~\forall n$, thus the two bands are decoupled, resulting in $E_{\text{Trivial},\text{LL},n}=(2n+1)/m_1l_B^2$ and $(2n+1)/m_2l_B^2$ instead. Thus, according to Eq.~\eqref{RGN}, $\partial_\ep M$ and $\partial_\ep \chi$ both vanish in a trivial case without nontrivial quantum geometry.  

\subsection{Magnetic properties of Euler bands}
Using Eq.~\eqref{E_n} we solve for $n$ in terms of $\ep=E_{\text{Eu},\text{LL},n,\pm}$ and obtain,
\begin{align}
    n &= \frac{1}{2} + \frac{(m_1+m_2) \ep\pm \sqrt{(m_1-m_2)^2\ep^2+4\a^2m_1m_2(3m_1+m_2)(3m_2+m_1)}}{8\a m_1m_2}
    \nonumber\\
    &=
    \frac{1}{2} + \frac{m_1+m_2 }{2B}\ep + \sqrt{\frac{\ep^2(m_1-m_2)^2}{4B^2}
    +\frac{(3m_1+m_2)(m_1+3m_2)}{16m_1m_2}}
    \\
    &=
    \left[\ep(m_1+m_2)\pm m_1m_2\sqrt{\frac{\ep^2(m_1-m_2)^2}{m_1^2m_2^2}}\right]\frac{1}{2B}
    +\frac{1}{2}
    \pm\sqrt{\frac{m_1^2m_2^2}{\ep^2(m_1-m_2)^2}}
    \frac{(3m_1+m_2)(3m_2+m_1)}{16}B
    + \bigO[B^2]\;.
\end{align}
For simplicity, in this expression, we have set flux quantum $\phi_0=h/e=1$ and $\hbar = 1$. For the case of interest, namely $m_1,m_2<0$, the pair of Euler bands would have equal and opposite contributions, resulting in $\partial_\epsilon\chi|_{\ep=\mu}=0$, thus resulting in constant magnetic susceptibility $\chi(\ep)$ around the node. One cannot obtain further information about the magnetic susceptibility from the Landau level, for example, whether the system is diamagnetic or paramagnetic. Instead, this has to be done using the Green's function method, as detailed in the next section. Crucially, and centrally to our work, this necessitates going significantly beyond the formulation of anomalous Landau levels and associated quantum geometry studied in previous works~\cite{Bouhon2019nonabelian, Rhim2020, Guan2022}.

\section{Further details on orbital magnetic susceptibility}\label{app::B}
\subsection{Fukuyama formula}

Within the equilibrium Matsubara formalism, the orbital magnetic susceptibility is given by\cite{Fukuyama1971}:
\beq{Fukuyama}
    \chi_\text{O} = \frac{e^2}{\hbar^2 c^2 }k_BT\sum_n\sum_\kv \Tr[j_x G j_y G j_x G j_y G]
\eec
where $G$ is the Matsubara Green's function at temperature $T$ and chemical potential $\mu$ that reads:
\begin{align}
    G^{-1} &= i\om_n + \mu - H \nonumber\\
    &= \sum_a (i\om_n+\mu-\ep_a)P_a\ , \\
    G &= \sum_a (i\om_n+\mu-\ep_a)^{-1}P_a\ .
\end{align}
Here, $\om_n = (2n+1)\pi k_B T$ are the Matsubara frequencies; $\ep_a$ and $P_a = \ket{u_{a\kv}} \bra{u_{a\kv}}$ are respectively the dispersion and band projector matrices for band $a$. We note that analogous calculations have been performed for centrosymmetric potentials in nontopological contexts \cite{Ogata2015}. However, this fails to describe systems beyond the $s$-wave orbital-based Lieb Hamiltonians. We define the velocity operator as $j_\mu = \partial_\mu H=\sum_a(\partial_\mu\ep_a P_a + \ep_a \partial_\mu P_a)$. As such, it follows that:
\begin{align}
    G j_\mu G &= G\partial_\mu H G
    \nonumber\\
    &= -G (\partial_\mu G^{-1}) G
    \nonumber\\
    &= GG^{-1}\partial_\mu G
    \nonumber\\
    &= \partial_\mu G\;,\label{GjG}
\end{align}
and we can rewrite the magnetic susceptibility as:
\beq{chi}
    \chi_\text{O}  = \frac{e^2}{\hbar^2 c^2}k_BT\sum_n\sum_\kv \Tr[j_x\partial_y G j_x\partial_y G]\;.
\eeq
We first consider the trivial case, such that $\partial_\mu P_a=0$, where we are left with:
\begin{align}
    \chi_\text{O}  &=  \frac{e^2}{\hbar^2 c^2}k_BT\sum_n\sum_\kv\sum_a \partial_x \ep_a\frac{\partial_y \ep_a}{(i\om_n+\mu-\ep_a)^2}\partial_x \ep_a\frac{\partial_y \ep_a}{(i\om_n+\mu-\ep_a)^2}
    \nonumber\\
    &=  \frac{e^2}{\hbar^2 c^2}\sum_a \frac{1}{3!}\int\frac{kdkd\th}{4\pi^2}(\partial_x\ep_a\partial_y\ep_a)^2\partial_{\ep_a}^3 f_a
    \nonumber\\
    &=  \frac{e^2}{\hbar^2 c^2}\sum_a \int\frac{kdkd\th}{4\pi^2}\frac{1}{3!}\frac{k_x^2k_y^2}{m_a^4}\partial_{\ep_a}^3 f_a
    \nonumber\\
    &=  \frac{e^2}{\hbar^2 c^2}\sum_a \int\frac{kdkd\th}{4\pi^2}\frac{1}{3!}\frac{k^4\sin^2\th\cos^2\th}{m_a^4}\partial_{\ep_a}^3 f_a
    \nonumber\\
    &=  \frac{e^2}{\hbar^2 c^2}\sum_a \int^{\ep_0}_0\frac{m_ad\ep }{4\pi^2}\frac{1}{3!}\frac{\pi}{4}\frac{4\ep^2}{m_a^2}\partial_{\ep_a}^3 f_a
    \nonumber\\
    &\ra -\frac{e^2}{12\pi\hbar^2 c^2}\sum_a \frac{1}{|m_a|}\; , \label{chiE0}
\end{align}
where $m_a$ is an effective mass,  $f_a = \big[\text{exp}[{(\ep_a-\mu)/T}]+1\big]^{-1}$ is the Fermi-Dirac distribution, the lattice constant is set to $a_0=1$, and we used the Matsubara sum Eq.~\eqref{MS}. We call this the energetic contribution $\chi_\text{E}$, which is identical, whether the nontrivial quantum geometry is present or not. Below, we will now discuss the other contributions, related to the quantum geometry of Bloch states.

\subsubsection{General decomposition}
To calculate the orbital susceptibility contribution when the geometry is nontrivial, we will begin by rewriting Eq.~\eqref{chi}, in~terms of the projector matrices ($P_a$) and band dispersions ($\ep_a$):
\begin{align}
    \chi_\text{O}  &= \frac{e^2}{\hbar^2 c^2}k_BT\sum_n\sum_\kv \sum_{a,b,c,d}\mathrm{Tr}\left\{
    (\partial_x \ep_a P_a+ \ep_a\partial_x P_a)
    \left[\frac{P_b\partial_y\ep_b}{(i\om_n+\mu-\ep_b)^2}+\frac{\partial_yP_b}{i\om_n+\mu-\ep_b}\right]\right.
    \nonumber\\
    &\quad\left.(\partial_x \ep_c P_c+ \ep_c\partial_x P_c)
    \left[\frac{P_d\partial_y\ep_d}{(i\om_n+\mu-\ep_d)^2}+\frac{\partial_yP_d}{i\om_n+\mu-\ep_d}\right]\right\}
    \nonumber\\
    &= \frac{e^2}{\hbar^2 c^2}k_BT\sum_n\sum_\kv \sum_{a,b,c,d} \frac{\partial_x\ep_a\partial_y\ep_b\partial_x\ep_c\partial_y\ep_d}
    {(i\om_n+\mu-\ep_b)^2(i\om_n+\mu-\ep_d)^2} 
    \Tr[P_aP_bP_cP_d]
    \nonumber\\
    &+ 
    2\frac{\ep_a\partial_y\ep_b\partial_x\ep_c\partial_y\ep_d}
    {(i\om_n+\mu-\ep_b)^2(i\om_n+\mu-\ep_d)^2} 
    \Tr[\partial_xP_aP_bP_cP_d]
    +
    2\frac{\partial_x\ep_a\partial_x\ep_c\partial_y\ep_d}
    {(i\om_n+\mu-\ep_b)(i\om_n+\mu-\ep_d)^2} 
    \Tr[P_a\partial_yP_bP_cP_d]
    \nonumber\\
    &+ 
    2\frac{\ep_a\partial_x\ep_c\partial_y\ep_d}
    {(i\om_n+\mu-\ep_b)(i\om_n+\mu-\ep_d)^2} 
    \Tr[\partial_xP_a\partial_yP_bP_cP_d]
    +
    2\frac{\ep_c\partial_x\ep_a\partial_y\ep_d}
    {(i\om_n+\mu-\ep_b)(i\om_n+\mu-\ep_d)^2} 
    \Tr[P_a\partial_yP_b\partial_xP_cP_d]
    \nonumber\\
    &+
    \frac{\ep_a\ep_c\partial_y\ep_b\partial_y\ep_d}
    {(i\om_n+\mu-\ep_b)^2(i\om_n+\mu-\ep_d)^2} 
    \Tr[\partial_xP_aP_b\partial_xP_cP_d]
    + 
    \frac{\partial_x\ep_a\partial_x\ep_c}
    {(i\om_n+\mu-\ep_b)(i\om_n+\mu-\ep_d)} 
    \Tr[P_a\partial_yP_bP_c\partial_yP_d]
    \nonumber\\
    &+2\frac{\ep_c\partial_x\ep_a}
    {(i\om_n+\mu-\ep_b)(i\om_n+\mu-\ep_d)} 
    \Tr[P_a\partial_yP_b\partial_xP_c\partial_yP_d]
    + 
    2\frac{\ep_a\ep_c\partial_y\ep_d}
    {(i\om_n+\mu-\ep_b)(i\om_n+\mu-\ep_d)^2} 
    \Tr[\partial_xP_a\partial_yP_b\partial_xP_cP_d]
    \nonumber\\
    &+
    \frac{\ep_a\ep_c}
    {(i\om_n+\mu-\ep_b)(i\om_n+\mu-\ep_d)} 
    \Tr[\partial_xP_a\partial_yP_b\partial_xP_c\partial_yP_d]\ .
    \nonumber\\
\end{align}
Below, we will refer to each term as $\chi_1,\chi_2\ldots, \chi_{10}$. To simplify further, we first employ the sum rules of projector matrices and their derivatives:
\begin{align}
    \sum_{a,b,c,d} \Tr[P_aP_bP_cP_d]
    &=
    \sum_{a,b,c,d} \Tr[P_a]\d_{ab}\d_{ac}\d_{ad}
    \nonumber\\
    &=
    \sum_{a,b,c,d} \d_{ab}\d_{ac}\d_{ad}\ ,
    \\
    \sum_{a,b,c,d} \Tr[\partial_\mu P_aP_bP_cP_d]
    &=
    \sum_{a,b,c,d} \Tr[\partial_\mu P_aP_b]\d_{bc}\d_{bd}
    \nonumber\\
    &=
    \sum_{a,b,c,d} \Tr[\ket{\partial_\mu a}\braket{a}{b}\bra{b}+
    \ket{a}\braket{\partial_\mu a}{b}\bra{b}]\d_{bc}\d_{bd}
    \nonumber\\
    &=
    \sum_{a,b,c,d} \Tr[\braket{b}{\partial_\mu a}+\braket{\partial_\mu a}{b}]\d_{ab}\d_{bc}\d_{bd}
    \nonumber\\
    &=
    \sum_{a,b,c,d} \Tr[\partial_\mu\braket{a}{a}]\d_{ab}\d_{ac}\d_{ad}
    \nonumber\\
    &= 0\ ,
    \label{Tr1}
    \\
    \sum_{a,b,c,d} \Tr[\partial_\mu P_a\partial_\nu P_bP_cP_d]
    &=
    \sum_{a,b,c,d} \Tr[\partial_\mu P_a\partial_\nu P_bP_c]\d_{cd}
    \nonumber\\
    &=
    \sum_{a,b,c,d} \Tr[
    \ket{\partial_\mu a}\braket{a}{\partial_\nu b}\bra{b}P_c
    +\ket{\partial_\mu a}\braket{a}{b}\bra{\partial_\nu b}P_c
    +\ket{ a}\braket{\partial_\mu a}{\partial_\nu b}\bra{b}P_c
    +\ket{ a}\braket{\partial_\mu a}{ b}\bra{\partial_\nu b}P_c
    ]\d_{cd}
    \nonumber\\
    &=
    \sum_{a,b,c,d} 
    \lb\braket{b}{\partial_\mu a}\braket{a}{\partial_\nu b}\d_{bc}
    +\braket{c}{\partial_\mu a}\braket{\partial_\nu a}{c}\d_{ab}
    +\braket{\partial_\mu a}{\partial_\nu a}\d_{bc}\d_{ac}
    +\braket{\partial_\mu a}{ b}\braket{\partial_\nu b}{a}\d_{ac}
    \rb\d_{cd}
    \nonumber\\
    &=
    \sum_{a,b,c,d} 
    \lb-\xi_\mu^{ba}\xi_\nu^{ab}\d_{bc}
    +\xi_\mu^{ca}\xi_\nu^{ac}\d_{ab}
    +\braket{\partial_\mu a}{\partial_\nu a}\d_{bc}\d_{ac}
    -\xi_\mu^{ab}\xi_\nu^{ba}\d_{ac}
    \rb\d_{cd}\; ,
    \label{PmuPnu}
    \end{align}
    \begin{align}
    &\sum_{a,b,c,d} \Tr[\partial_\mu P_a\partial_\nu P_bP_cP_d] + 
    \sum_{a,b,c,d} \Tr[P_c\partial_\nu P_b\partial_\mu P_aP_d]
    \nonumber\\
    &= \sum_{a,b,c,d} \Tr[(\partial_\mu P_a\partial_\nu P_b+\partial_\nu P_b\partial_\mu P_a)P_c]\d_{cd}
    \nonumber\\
    &= \sum_{a,b,c,d} 
    \lb-\xi_\mu^{ba}\xi_\nu^{ab}\d_{bc}
    +\xi_\mu^{ca}\xi_\nu^{ac}\d_{ab}
    +\braket{\partial_\mu a}{\partial_\nu a}\d_{bc}\d_{ac}
    -\xi_\mu^{ab}\xi_\nu^{ba}\d_{ac}
    -\xi_\nu^{ab}\xi_\mu^{ba}\d_{ac}
    +\xi_\nu^{cb}\xi_\mu^{bc}\d_{ab}
    +\braket{\partial_\nu a}{\partial_\mu a}\d_{bc}\d_{ac}
    -\xi_\nu^{ba}\xi_\mu^{ab}\d_{bc}
    \rb\d_{cd}
    \nonumber\\
    &=
    2\sum_{a,b,c,d} 
    \lb-\Re\Big[\xi_\mu^{ab}\xi_\nu^{ba}\Big]\d_{bc}
    +\Re\Big[\xi_\mu^{bc}\xi_\nu^{cb}\Big](\d_{ab}-\d_{ac})
    +\Re\Big[\braket{\partial_\mu a}{\partial_\nu a}\Big]\d_{bc}\d_{ac}
    \rb\d_{cd}
    \nonumber\\
    &=
    2\sum_{a,b,c,d} 
    \lb-g_{\mu\nu}^{ab}\d_{bc}
    +g_{\mu\nu}^{bc}(\d_{ab}-\d_{ac})
    +\Re\Big[\braket{\partial_\mu a}{\partial_\nu a}\Big]\d_{bc}\d_{ac}
    \rb\d_{cd}
    \ ,\label{Tr2_1}
    \\
    &\sum_{a,b,c,d} \Tr[\partial_\mu P_a P_b\partial_\mu P_cP_d]
    \nonumber\\
    &=
    \sum_{a,b,c,d} \Tr[\partial_\mu P_a \partial_\mu(P_b P_c)P_d-\partial_\mu P_a \partial_\mu P_b P_cP_d]
    \nonumber\\
    &=
    \sum_{a,b,c,d} \Tr[\partial_\mu P_a \partial_\mu P_b\d_{bc}P_d-\partial_\mu P_a \partial_\mu P_b\d_{cd} P_d]
    \nonumber\\
    &=
    \sum_{a,b,c,d} \Tr[\partial_\mu P_a \partial_\mu P_bP_d(\d_{bc}-\d_{cd})]
    \nonumber\\
    &= 2\sum_{a,b,c,d} \Re\Big[\xi_\mu^{db}\xi_\mu^{bd}\Big](\d_{bc}-\d_{cd})\d_{ab}
    \nonumber\\
    &= 2\sum_{a,b,c,d} g_{\mu \mu}^{bd}(\d_{bc}-\d_{cd})\d_{ab}
    \ ,\label{Tr2_2}
    \end{align}
    \begin{align}
    &\sum_{a,b,c,d} \Tr[P_a\partial_\mu P_b\partial_\nu P_c \partial_\mu P_d]
    \nonumber\\
    &=\sum_{a,b,c,d} \braket{d}{a}\braket{a}{\partial_\mu b}\braket{b}{\partial_\nu c}\braket{c}{\partial_\mu d} + 
    \braket{\partial_\mu d}{a}\braket{a}{\partial_\mu b}\braket{b}{\partial_\nu c}\braket{c}{d} +
    \braket{d}{a}\braket{a}{\partial_\mu b}\braket{b}{c}\braket{\partial_\nu c}{\partial_\mu d} + 
    \braket{\partial_\mu d}{a}\braket{a}{\partial_\mu b}\braket{b}{c}\braket{\partial_\nu c}{d}
    \nonumber\\
    &+
    \braket{d}{a}\braket{a}{b}\braket{\partial_\mu b}{\partial_\nu c}\braket{c}{\partial_\mu d} + 
    \braket{\partial_\mu d}{a}\braket{a}{b}\braket{\partial_\mu b}{\partial_\nu c}\braket{c}{d} +
    \braket{d}{a}\braket{a}{b}\braket{\partial_\mu b}{c}\braket{\partial_\nu c}{\partial_\mu d} + 
    \braket{\partial_\mu d}{a}\braket{a}{b}\braket{\partial_\mu b}{c}\braket{\partial_\nu c}{d}
    \nonumber\\
    &=
    2\sum_{a,b,c,d}
    \d_{ab}\d_{ad}
    \Re\Big[\braket{\partial_\mu a}{c}\braket{\partial_\nu c}{\partial_\mu a}\Big] 
    + \d_{ad}\d_{bc}
    \Re\Big[\braket{a}{\partial_\mu c}\braket{\partial_\nu c}{\partial_\mu a}\Big]
    \nonumber\\
    &+\d_{cd}
    \Re\Big[\braket{a}{\partial_\mu b}\braket{b}{\partial_\nu c}\braket{\partial_\mu c}{a}\Big]
    +\d_{ad}
    \Re\Big[\braket{a}{\partial_\mu b}\braket{b}{\partial_\nu c}\braket{c}{\partial_\mu a}\Big]
    \nonumber\\
    &=
    2\sum_{a,b,c,d}
    \d_{ab}(\d_{dc}-\d_{ad})
    \Re\Big[\braket{a}{\partial_\mu c}\braket{\partial_\nu c}{\partial_\mu a}\Big] + (\d_{ad}-\d_{cd})
    \Re\Big[\braket{a}{\partial_\mu b}\braket{b}{\partial_\nu c}\braket{c}{\partial_\mu a}\Big]
    \nonumber\\
    &=
    2\sum_{a,b,c,d}(\d_{ad}-\d_{cd})
    \Re\Big[\braket{a}{\partial_\mu b}\braket{b}{\partial_\nu c}\braket{c}{\partial_\mu a}-\d_{ab}\braket{a}{\partial_\mu c}\braket{\partial_\nu c}{\partial_\mu a}\Big]\ ,
    \label{Tr3}
    \\
    &\sum_{a,b,c,d}\Tr[\partial_\mu P_a\partial_\nu P_b\partial_\mu P_c\partial_\nu P_d]
    \nonumber\\
    &=2\sum_{a,b,c,d}\Re\Big[
    \braket{a}{\partial_\nu b}\braket{b}{\partial_\mu c}\braket{c}{\partial_\nu d}\braket{d}{\partial_\mu a}
    +
    \braket{\partial_\mu a}{\partial_\nu c}\braket{\partial_\mu c}{\partial_\nu a} \d_{ab}\d_{cd}
    \nonumber\\
    &+
    2\d_{cd}(
    \braket{a}{\partial_\nu b}\braket{b}{\partial_\mu c}\braket{\partial_\nu c}{\partial_\mu a}
    -
    \braket{\partial_\mu a}{\partial_\nu b}\braket{b}{\partial_\mu c}\braket{c}{\partial_\nu a}
    +
    \braket{a}{\partial_\mu b}\braket{\partial_\nu b}{\partial_\mu c}\braket{c}{\partial_\nu a}
    )\Big]
    \nonumber\\
    &=2\sum_{a,b,c,d}\Re\Big[
    \braket{a}{\partial_\nu b}\braket{b}{\partial_\mu c}\braket{c}{\partial_\nu d}\braket{d}{\partial_\mu a}
    +
    \braket{\partial_\mu a}{\partial_\nu c}\braket{\partial_\mu c}{\partial_\nu a} \d_{ab}\d_{cd}
    \nonumber\\
    &+
    2\d_{cd}(
    \braket{a}{\partial_\nu b}\braket{b}{\partial_\mu c}\braket{\partial_\nu c}{\partial_\mu a}
    -
    \braket{\partial_\mu a}{\partial_\nu b}\braket{b}{\partial_\mu c}\braket{c}{\partial_\nu a}
    +
    \braket{a}{\partial_\mu b}\braket{\partial_\nu b}{\partial_\mu d}\braket{d}{\partial_\nu a}
    )\Big]
    \; ,
    \label{Tr4}
\end{align}

where we have an additional constraint of only being able to swap index $a\leftrightarrow c$ and $b\leftrightarrow d$ under the certain condition, due to the symmetry of prefactor that encapsulate the contribution from dispersion. For instance, to be able to swap $a$ and $c$, the corresponding projector matrix has to be acted upon (or not acted upon) with derivatives simultaneously. Note that here we defined the interband Berry connection $ \xi^{ab}_\mu=i\braket{a}{\partial_\mu b}$, using a shorthand notation $\ket{a} \equiv \ket{u_{a\kv}}$. Thus, the interband metric $g_{\mu\nu}^{ab}$, with $a\neq b$, amounts to:

\begin{align}
    g_{\mu\nu}^{ab} &= \Re\Big[\xi_\mu^{ab}\xi_\nu^{ba}\Big]
    \nonumber\\
    &= -\Re\Big[\braket{a}{\partial_\mu b}\braket{b}{\partial_\nu a}\Big]
    \nonumber\\
    &=\frac{\Re\Big[j_\mu^{ab}j_{\nu}^{ba}\Big]}{\ep_{ab}^2}\;,
    \label{metric}
    \\
    j_\mu^{ab} &= \bra{a}j_\mu\ket{b}
    \nonumber\\
    &= \bra{a}\sum_c (\partial_\mu\ep_c\ket{c}\bra{c}
    +\ep_c\ket{\partial_\mu c}\bra{c}
    +\ep_c\ket{c}\bra{\partial_\mu c})\ket{b}
    \nonumber\\
    &=\partial_\mu \ep_a\d_{ab}+ \ep_b\braket{a}{\partial_\mu b}+\ep_a\braket{\partial_\mu a}{b}
    \nonumber\\
    &= \partial_\mu \ep_a\d_{ab}- \ep_{ab}\braket{a}{\partial_\mu b}\;,
\end{align}
where $\ep_{ab}=\ep_a-\ep_b$. Using properties of projector matrices derived above, we can simplify the various terms of $\chi_\text{O}$ accordingly. We begin by calculating $\chi_1$, which is the purely energetic contribution:
\begin{align}
    \chi_1
    &=
    \frac{e^2}{\hbar^2 c^2}k_BT\sum_n\sum_\kv \sum_{a,b,c,d} \frac{\partial_x\ep_a\partial_y\ep_b\partial_x\ep_c\partial_y\ep_d}
    {(i\om_n+\mu-\ep_b)^2(i\om_n+\mu-\ep_d)^2} 
    \Tr[P_aP_bP_cP_d]
    \nonumber\\
    &=
    \frac{e^2}{\hbar^2 c^2}k_BT\sum_n\sum_\kv \sum_{a} \frac{\partial_x\ep_a\partial_y\ep_a\partial_x\ep_a\partial_y\ep_a}
    {(i\om_n+\mu-\ep_a)^4} 
    \nonumber\\
    &= \chi_\text{E}\;.
\end{align}
We then note $\chi_2 = \chi_3 = 0$ due to the vanishing of trace, according to Eq.~\eqref{Tr1}. For $\chi_4,\ \chi_5,\ \chi_6,\ \chi_7$, these are the terms involving only two bands upon applying sum rules, and can be related to the multiband metric. To begin with, $\chi_4$ and $\chi_5$ can be recombined into a contribution proportional to $g^{ab}_{xy}$:
\begin{align}
    \chi_4 + \chi_5 &= 
    2\frac{e^2}{\hbar^2 c^2}k_BT\sum_n\sum_\kv \sum_{a,b,c,d}\frac{\ep_a\partial_x\ep_c\partial_y\ep_d}
    {(i\om_n+\mu-\ep_b)(i\om_n+\mu-\ep_d)^2} 
    \lb\Tr[\partial_xP_a\partial_yP_bP_cP_d]
    +
    \Tr[P_c\partial_yP_b\partial_xP_aP_d]\rb
    \nonumber\\
    &= 
    4\frac{e^2}{\hbar^2 c^2}k_BT\sum_n\sum_\kv \sum_{a,b,c,d}\frac{\ep_a\partial_x\ep_c\partial_y\ep_d}
    {(i\om_n+\mu-\ep_b)(i\om_n+\mu-\ep_d)^2} 
    \lb-g_{xy}^{ab}\d_{bc}
    +g_{xy}^{bc}(\d_{ab}-\d_{ac})
    +\Re\Big[\braket{\partial_x a}{\partial_y a}\Big]\d_{bc}\d_{ac}
    \rb\d_{cd}
    \nonumber\\
    &= \chi_4'+\chi_5'\ ,
    \nonumber\\
    \chi_4'&=
    4\frac{e^2}{\hbar^2 c^2}k_BT\sum_n\sum_\kv \sum_{a,b}\frac{\ep_a\partial_x\ep_b\partial_y\ep_b}
    {(i\om_n+\mu-\ep_b)^3} 
    \lb-g_{xy}^{ab}
    +\Re\Big[\braket{\partial_x a}{\partial_y a}\Big]\d_{ab}
    \rb
    \nonumber\\
    &=4\frac{e^2}{\hbar^2 c^2}k_BT\sum_n\sum_\kv \sum_{a,b}\frac{\partial_x\ep_b\partial_y\ep_b}
    {(i\om_n+\mu-\ep_b)^3}
    \lb
    -\ep_ag_{xy}^{ab}
    +
    \ep_b\Re\Big[\braket{\partial_x b}{a}\braket{a}{\partial_y b}\Big]
    \rb 
    \nonumber\\
    &=4\frac{e^2}{\hbar^2 c^2}k_BT\sum_n\sum_\kv \sum_{a,b}\frac{\partial_x\ep_b\partial_y\ep_b}
    {(i\om_n+\mu-\ep_b)^3}
    \ep_{ba}g_{xy}^{ab}\ ,
    \nonumber\\
    \chi_5'
    &=
    4\frac{e^2}{\hbar^2 c^2}k_BT\sum_n\sum_\kv \sum_{a,b,c,d}\frac{\ep_a\partial_x\ep_c\partial_y\ep_d}
    {(i\om_n+\mu-\ep_b)(i\om_n+\mu-\ep_d)^2} 
    g_{xy}^{bc}(\d_{ab}-\d_{ac})
    \d_{cd}
    \nonumber\\
    &=
    -4\frac{e^2}{\hbar^2 c^2}k_BT\sum_n\sum_\kv \sum_{b,c}\frac{\partial_x\ep_c\partial_y\ep_c}
    {(i\om_n+\mu-\ep_b)(i\om_n+\mu-\ep_c)^2} \ep_{cb}
    g_{xy}^{bc}
    \nonumber\\
    &=
    -4\frac{e^2}{\hbar^2 c^2}k_BT\sum_n\sum_\kv \sum_{a,b}\frac{\partial_x\ep_b\partial_y\ep_b}
    {(i\om_n+\mu-\ep_a)(i\om_n+\mu-\ep_b)^2} \ep_{ba}
    g_{xy}^{ab}\ ,
    \nonumber\\
    \chi_4+\chi_5 &=
    4\frac{e^2}{\hbar^2 c^2}k_BT\sum_n\sum_\kv \sum_{a,b}\frac{\ep_{ba}^2\partial_x\ep_b\partial_y\ep_b}
    {(i\om_n+\mu-\ep_b)^3(i\om_n+\mu-\ep_a)}
    g_{xy}^{ab}\ .
    \end{align}
    Similarly, $\chi_6$ can be shown to depend on $g^{ab}_{xx}$:
    \begin{align}
    \chi_6 &=
    \frac{e^2}{\hbar^2 c^2}k_BT\sum_n\sum_\kv\sum_{a,b,c,d}
    \frac{\ep_a\ep_c\partial_y\ep_b\partial_y\ep_d}
    {(i\om_n+\mu-\ep_b)^2(i\om_n+\mu-\ep_d)^2} 
    \Tr[\partial_xP_aP_b\partial_xP_cP_d]
    \nonumber\\
    &=
    2\frac{e^2}{\hbar^2 c^2}k_BT\sum_n\sum_\kv\sum_{a,b,c,d} 
    \frac{\ep_a\ep_c\partial_y\ep_b\partial_y\ep_d}
    {(i\om_n+\mu-\ep_b)^2(i\om_n+\mu-\ep_d)^2} 
    g_{xx}^{bd}(\d_{bc}-\d_{cd})\d_{ab}
    \nonumber\\
    &=
    2\frac{e^2}{\hbar^2 c^2}k_BT\sum_n\sum_\kv\sum_{b,d} 
    \frac{\ep_b\partial_y\ep_b\partial_y\ep_d}
    {(i\om_n+\mu-\ep_b)^2(i\om_n+\mu-\ep_d)^2} \ep_{bd}g_{xx}^{bd}
    \nonumber\\
    &=
    -2\frac{e^2}{\hbar^2 c^2}k_BT\sum_n\sum_\kv\sum_{a,b} 
    \frac{\partial_y\ep_a\partial_y\ep_b}
    {(i\om_n+\mu-\ep_a)^2(i\om_n+\mu-\ep_b)^2} \ep_b\ep_{ab}g_{xx}^{ba}
    \nonumber\\
    &=
    \frac{e^2}{\hbar^2 c^2}k_BT\sum_n\sum_\kv\sum_{a,b} 
    \frac{\partial_y\ep_a\partial_y\ep_b}
    {(i\om_n+\mu-\ep_a)^2(i\om_n+\mu-\ep_b)^2} \ep_{ab}^2g_{xx}^{ba}\; .
    \end{align}
For $\chi_7$, which depends on $g^{ab}_{yy}$, the term can be rewritten after some algebra, to obtain similar functional form as $\chi_6$:
    \begin{align}
    \chi_7 &=
    \frac{e^2}{\hbar^2 c^2}k_BT\sum_n\sum_\kv\sum_{a,b,c,d}
    \frac{\partial_x\ep_a\partial_x\ep_c}
    {(i\om_n+\mu-\ep_b)(i\om_n+\mu-\ep_d)} 
    \Tr[P_a\partial_yP_bP_c\partial_yP_d]
    \nonumber\\
    &=
    \frac{e^2}{\hbar^2 c^2}k_BT\sum_n\sum_\kv\sum_{a,b,c,d}
    \frac{\partial_x\ep_a\partial_x\ep_c}
    {(i\om_n+\mu-\ep_b)(i\om_n+\mu-\ep_d)} 
    \Tr[\partial_yP_bP_c\partial_yP_dP_a]
    \nonumber\\
    &=
    2\frac{e^2}{\hbar^2 c^2}k_BT\sum_n\sum_\kv\sum_{a,b,c,d}
    \frac{\partial_x\ep_a\partial_x\ep_c}
    {(i\om_n+\mu-\ep_b)(i\om_n+\mu-\ep_d)} 
    g_{yy}^{ca}(\d_{cd}-\d_{ad})\d_{bc}
    \nonumber\\
    &=
    2\frac{e^2}{\hbar^2 c^2}k_BT\sum_n\sum_\kv\sum_{a,b,d}
    \frac{\partial_x\ep_a\partial_x\ep_b}
    {(i\om_n+\mu-\ep_b)(i\om_n+\mu-\ep_d)} 
    g_{yy}^{ba}(\d_{bd}-\d_{ad})
    \nonumber\\
    &=
    2\frac{e^2}{\hbar^2 c^2}k_BT\sum_n\sum_\kv\lb
    \sum_{a,b}
    \frac{\partial_x\ep_a\partial_x\ep_b}
    {(i\om_n+\mu-\ep_b)^2} 
    g_{yy}^{ba}
    -
    \sum_{b,d}
    \frac{\partial_x\ep_d\partial_x\ep_b}
    {(i\om_n+\mu-\ep_b)(i\om_n+\mu-\ep_d)} 
    g_{yy}^{bd}
    \rb
    \nonumber\\
    &=
    -2\frac{e^2}{\hbar^2 c^2}k_BT\sum_n\sum_\kv
    \sum_{a,b}
    \frac{\partial_x\ep_a\partial_x\ep_b}
    {(i\om_n+\mu-\ep_a)(i\om_n+\mu-\ep_b)^2} \ep_{ab}
    g_{yy}^{ba}
    \nonumber\\
    &=
    \frac{e^2}{\hbar^2 c^2}k_BT\sum_n\sum_\kv
    \sum_{a,b}
    \frac{\partial_x\ep_a\partial_x\ep_b}
    {(i\om_n+\mu-\ep_a)^2(i\om_n+\mu-\ep_b)^2} \ep_{ab}^2
    g_{yy}^{ba}\; .
\end{align}
We then compute $\chi_8$ and $\chi_9$ using Eq.~\eqref{Tr3}, which we find to correspond to transitions between three distinct bands, thus are vanishing in two-band limits:
\begin{align}
    \chi_8
    &= 
    2\frac{e^2}{\hbar^2c^2}k_BT\sum_n\sum_\kv\sum_{a,b,c,d}
    \frac{\ep_c\partial_x\ep_a}
    {(i\om_n+\mu-\ep_b)(i\om_n+\mu-\ep_d)} 
    \Tr[P_a\partial_yP_b\partial_xP_c\partial_yP_d]
    \nonumber\\
    &= 
    4\frac{e^2}{\hbar^2c^2}k_BT\sum_n\sum_\kv\sum_{a,b,c,d}
    \frac{\ep_c\partial_x\ep_a}
    {(i\om_n+\mu-\ep_b)(i\om_n+\mu-\ep_d)} 
    (\d_{ad}-\d_{cd})
    \Re\Big[\braket{a}{\partial_y b}\braket{b}{\partial_x c}\braket{c}{\partial_y a}-\d_{ab}\braket{a}{\partial_y c}\braket{\partial_x c}{\partial_y a}\Big]
    \nonumber\\
    &=
    4\frac{e^2}{\hbar^2c^2}k_BT\sum_n\sum_\kv\sum_{a,b,c,d}
    \frac{\ep_{ba}\ep_c\partial_x\ep_a}
    {(i\om_n+\mu-\ep_a)(i\om_n+\mu-\ep_b)(i\om_n+\mu-\ep_d)} 
    (\d_{ad}-\d_{cd})
    \Re\Big[\braket{a}{\partial_y b}\braket{b}{\partial_x c}\braket{c}{\partial_y a}\Big]
    \nonumber\\
    &=
    4\frac{e^2}{\hbar^2c^2}k_BT\sum_n\sum_\kv\sum_{a,b,c}
    \frac{\ep_c\ep_{ba}\ep_{ac}\partial_x\ep_a}
    {(i\om_n+\mu-\ep_a)^2(i\om_n+\mu-\ep_b)(i\om_n+\mu-\ep_c)} 
    \Re\Big[\braket{a}{\partial_y b}\braket{b}{\partial_x c}\braket{c}{\partial_y a}\Big]
    \nonumber\\
    &=
    -2\frac{e^2}{\hbar^2c^2}k_BT\sum_n\sum_\kv\sum_{a,b,c}
    \frac{\ep_{ab}\ep_{bc}\ep_{ca}\partial_x\ep_a}
    {(i\om_n+\mu-\ep_a)^2(i\om_n+\mu-\ep_b)(i\om_n+\mu-\ep_c)} 
    \Re\Big[\braket{a}{\partial_y b}\braket{b}{\partial_x c}\braket{c}{\partial_y a}\Big]
    \ ,
    \nonumber
\end{align}
\begin{align}
    \chi_9
    &= 
    2\frac{e^2}{\hbar^2c^2}k_BT\sum_n\sum_\kv\sum_{a,b,c,d}
    \frac{\ep_a\ep_c\partial_y\ep_d}
    {(i\om_n+\mu-\ep_b)(i\om_n+\mu-\ep_d)^2} 
    \Tr[\partial_xP_a\partial_yP_b\partial_xP_cP_d]
    \nonumber\\
    &= 
    2\frac{e^2}{\hbar^2c^2}k_BT\sum_n\sum_\kv\sum_{a,b,c,d}
    \frac{\ep_b\ep_d\partial_y\ep_a}
    {(i\om_n+\mu-\ep_c)(i\om_n+\mu-\ep_a)^2} 
    \Tr[P_a\partial_xP_b\partial_yP_c\partial_xP_d]
    \nonumber\\
    &=
    4\frac{e^2}{\hbar^2c^2}k_BT\sum_n\sum_\kv\sum_{a,b,c,d}
    \frac{\ep_b\ep_d\partial_y\ep_a}
    {(i\om_n+\mu-\ep_c)(i\om_n+\mu-\ep_a)^2} 
    (\d_{ad}-\d_{cd})
    \Re\Big[\braket{a}{\partial_x b}\braket{b}{\partial_y c}\braket{c}{\partial_x a}-\d_{ab}\braket{a}{\partial_y c}\braket{\partial_x c}{\partial_y a}\Big]
    \nonumber\\
    &=
    4\frac{e^2}{\hbar^2c^2}k_BT\sum_n\sum_\kv\sum_{a,b,c,d}
    \frac{\ep_{ba}\ep_d\partial_y\ep_a}
    {(i\om_n+\mu-\ep_c)(i\om_n+\mu-\ep_a)^2} 
    (\d_{ad}-\d_{cd})
    \Re\Big[\braket{a}{\partial_x b}\braket{b}{\partial_y c}\braket{c}{\partial_x a}\Big]
    \nonumber\\
    &=
    4\frac{e^2}{\hbar^2c^2}k_BT\sum_n\sum_\kv\sum_{a,b,c}
    \frac{\ep_{ba}\ep_{ac}\partial_y\ep_a}
    {(i\om_n+\mu-\ep_a)^2(i\om_n+\mu-\ep_c)}
    \Re\Big[\braket{a}{\partial_x b}\braket{b}{\partial_y c}\braket{c}{\partial_x a}\Big]
    \nonumber\\
    &=
    -2\frac{e^2}{\hbar^2c^2}k_BT\sum_n\sum_\kv\sum_{a,b,c}
    \frac{\ep_{ab}\ep_{bc}\ep_{ca}\partial_y\ep_a}
    {(i\om_n+\mu-\ep_a)^2(i\om_n+\mu-\ep_b)(i\om_n+\mu-\ep_c)} 
    \Re\Big[\braket{a}{\partial_x b}\braket{b}{\partial_y c}\braket{c}{\partial_x a}\Big]
    \ ,
    \nonumber\\
    \chi_8 + \chi_9
    &=
    4\frac{e^2}{\hbar^2c^2}k_BT\sum_n\sum_\kv\sum_{a,b,c}
    \frac{\ep_{ba}\ep_{ac}\partial_y\ep_a}
    {(i\om_n+\mu-\ep_a)^2(i\om_n+\mu-\ep_c)}
    \Re\Big[\braket{a}{\partial_x b}\braket{b}{\partial_y c}\braket{c}{\partial_x a}\Big]
    + (x\leftrightarrow y)
    \ ,
\end{align} 
where we have used a symmetry between $b$ and $c$ in the contraction between Bloch states to rewrite $\chi_8$ and $\chi_9$ in a combined form.
Finally, we compute $\chi_{10}$:
\begin{align}
    \chi_{10}
    &=
    \frac{e^2}{\hbar^2c^2}k_BT\sum_n\sum_\kv\sum_{a,b,c,d}\frac{\ep_a\ep_c}
    {(i\om_n+\mu-\ep_b)(i\om_n+\mu-\ep_d)} 
    \Tr[\partial_xP_a\partial_yP_b\partial_xP_c\partial_yP_d]
    \nonumber\\
    &=
    2\frac{e^2}{\hbar^2c^2}k_BT\sum_n\sum_\kv\sum_{a,b,c,d}
    \frac{\ep_a\ep_c}
    {(i\om_n+\mu-\ep_b)(i\om_n+\mu-\ep_d)} 
    \Re\Big[
    \braket{a}{\partial_y b}\braket{b}{\partial_x c}\braket{c}{\partial_y d}\braket{d}{\partial_x a}
    +
    \braket{\partial_x a}{\partial_y c}\braket{\partial_x c}{\partial_y a} \d_{ab}\d_{cd}
    \nonumber\\
    &+
    2\d_{cd}
    \braket{a}{\partial_y b}\braket{b}{\partial_x c}\braket{\partial_y c}{\partial_x a}
    +
    2\d_{cd}
    \braket{a}{\partial_x b}\braket{\partial_y b}{\partial_x d}\braket{d}{\partial_y a}
    -
    2\d_{cd}
    \braket{\partial_x a}{\partial_y b}\braket{b}{\partial_x c}\braket{c}{\partial_y a}
    \Big]
    \nonumber\\
    &=
    2\frac{e^2}{\hbar^2c^2}k_BT\sum_n\sum_\kv\sum_{a,b,c,d} \Re\Big[
    \braket{a}{\partial_y b}\braket{b}{\partial_x c}\braket{c}{\partial_y d}\braket{d}{\partial_x a}\Big]
    \left(
    \frac{\ep_a\ep_c}
    {(i\om_n+\mu-\ep_b)(i\om_n+\mu-\ep_d)} 
    +
    \frac{\ep_a\ep_c}
    {(i\om_n+\mu-\ep_a)(i\om_n+\mu-\ep_c)}
    \right.
    \nonumber\\
    &
    \left.
    -
    \frac{2\ep_a\ep_c}
    {(i\om_n+\mu-\ep_b)(i\om_n+\mu-\ep_c)} 
    -
    \frac{2\ep_a\ep_b}
    {(i\om_n+\mu-\ep_a)(i\om_n+\mu-\ep_c)} 
    +
    \frac{2\ep_a\ep_b}
    {(i\om_n+\mu-\ep_b)(i\om_n+\mu-\ep_c)} 
    \right)
    \nonumber\\
    &=
    2\frac{e^2}{\hbar^2c^2}k_BT\sum_n\sum_\kv\sum_{a,b,c,d} \Re\Big[
    \braket{a}{\partial_y b}\braket{b}{\partial_x c}\braket{c}{\partial_y d}\braket{d}{\partial_x a}\Big]
    \nonumber\\
    &\lb
    \frac{-2\ep_a\ep_b\ep_{ab}}
    {(i\om_n+\mu-\ep_a)(i\om_n+\mu-\ep_b)(i\om_n+\mu-\ep_c)}
    +
    \frac{\ep_a\ep_c\ep_{ab}\ep_{cd}}
    {(i\om_n+\mu-\ep_a)(i\om_n+\mu-\ep_b)(i\om_n+\mu-\ep_c)(i\om_n+\mu-\ep_d)}
    \rb
    \nonumber\\
    &=
    \frac{e^2}{\hbar^2c^2}k_BT\sum_n\sum_\kv\sum_{a,b,c,d} \Re\Big[
    \braket{a}{\partial_y b}\braket{b}{\partial_x c}\braket{c}{\partial_y d}\braket{d}{\partial_x a}\Big]
    \frac{\ep_{ab}\ep_{bc}\ep_{cd}\ep_{da}}
    {(i\om_n+\mu-\ep_a)(i\om_n+\mu-\ep_b)(i\om_n+\mu-\ep_c)(i\om_n+\mu-\ep_d)}\ .
    \label{chi10}
\end{align}
To conclude, we have:
\begin{align}
    \chi_\text{O}  &= \chi_\text{E} + \chi_\geo+ \chi_\geo^{(1)} + \chi_\geo^{(2)}
    \ \,
    \\
    \chi_\text{E} &= \frac{e^2}{\hbar^2 c^2}k_BT\sum_n\sum_\kv \sum_{a} \frac{(\partial_x\ep_a\partial_y\ep_a)^2}
    {(i\om_n+\mu-\ep_a)^4} 
    \ \,
    \\
    \chi_{\text{geo}}
    &=
    \chi_{xy}+\chi_{xx}+\chi_{yy}
    \ \,
    \\
    \chi_{xy}
    &=
    \chi_4 + \chi_5
    \nonumber\\
    &=
    4\frac{e^2}{\hbar^2 c^2}k_BT\sum_n\sum_\kv \sum_{a,b}\frac{\ep_{ab}^2\partial_x\ep_b\partial_y\ep_b}
    {(i\om_n+\mu-\ep_b)^3(i\om_n+\mu-\ep_a)}
    g_{xy}^{ab}\ ,
\end{align}
\begin{align}
    \chi_{xx}
    &=
    \chi_6
    \nonumber\\
    &=
    \frac{e^2}{\hbar^2 c^2}k_BT\sum_n\sum_\kv
    \sum_{a,b}
    \frac{\ep_{ab}^2\partial_y\ep_a\partial_y\ep_b}
    {(i\om_n+\mu-\ep_a)^2(i\om_n+\mu-\ep_b)^2} g_{xx}^{ab} 
    \ ,
    \\
    \chi_{yy}
    &=
    \chi_7
    \nonumber\\
    &=
    \frac{e^2}{\hbar^2 c^2}k_BT\sum_n\sum_\kv
    \sum_{a,b}
    \frac{\ep_{ab}^2\partial_x\ep_a\partial_x\ep_b}
    {(i\om_n+\mu-\ep_a)^2(i\om_n+\mu-\ep_b)^2} g_{yy}^{ab} 
    \ ,
    \\
    \chi_\geo^{(1)}
    &=
    \chi_8 + \chi_9
    \nonumber\\
    &=
    4\frac{e^2}{\hbar^2c^2}k_BT\sum_n\sum_\kv\sum_{a,b,c}
    \frac{\ep_{ba}\ep_{ac}\partial_y\ep_a}
    {(i\om_n+\mu-\ep_a)^2(i\om_n+\mu-\ep_c)}
    \Re\Big[\braket{a}{\partial_x b}\braket{b}{\partial_y c}\braket{c}{\partial_x a}\Big]
    + (x\leftrightarrow y)
    \ ,
    \\
    \chi_\geo^{(2)}
    &= \chi_{10}
    \nonumber\\
    &=
    \frac{e^2}{\hbar^2c^2}k_BT\sum_n\sum_\kv\sum_{a,b,c,d} \Re\Big[
    \braket{a}{\partial_y b}\braket{b}{\partial_x c}\braket{c}{\partial_y d}\braket{d}{\partial_x a}\Big]
    \frac{\ep_{ab}\ep_{bc}\ep_{cd}\ep_{da}}
    {(i\om_n+\mu-\ep_a)(i\om_n+\mu-\ep_b)(i\om_n+\mu-\ep_c)(i\om_n+\mu-\ep_d)}
    \ .
\end{align}
\subsubsection{Two-band limit}
In the two-band limit, we note that $\chi^{(1)}_\geo$ is always vanishing. Interestingly, in the two-band limits with bands $a$ and $b$, $\chi_\geo^{(2)}$ is nonvanishing only if $a=c,\ b=d$:
\begin{align}
    \chi^{(2)}_\geo 
    &=
    \frac{e^2}{\hbar^2c^2}k_BT\sum_n\sum_\kv
    \Re\Big[
    (\braket{a}{\partial_y b}\braket{b}{\partial_x a})^2\Big]
    \frac{\ep_{ab}^4}
    {(i\om_n+\mu-\ep_a)^2(i\om_n+\mu-\ep_b)^2}
    \nonumber\\
    &=
    \frac{e^2}{\hbar^2c^2}k_BT\sum_n\sum_\kv \left[ (g_{xy}^{ab})^2 - \frac{(\Om_{xy}^{ab})^2}{4}\right]
    \frac{\ep_{ab}^4}
    {(i\om_n+\mu-\ep_a)^2(i\om_n+\mu-\ep_b)^2}
    \nonumber\\
    &=
    \frac{e^2}{\hbar^2c^2}\sum_\kv
    \left[ (g_{xy}^{ab})^2 - \frac{(\Om_{xy}^{ab})^2}{4}\right]
    {\ep_{ab}^2}\lb
    f'_a+f'_b
    -
    2\frac{f_{ab}}{\ep_{ab}}
    \rb,
    \label{chigeo2_2}
\end{align}
where $\Om_{xy}^{ab}=-2\Im[\braket{a}{\partial_y b}\braket{b}{\partial_x a}]$ is the multiband Berry curvature. We also compute the remaining nonvanishing terms, where we begin with the energetic term:
\begin{align}
    \chi_\text{E} &=
    \frac{e^2}{\hbar^2 c^2}\sum_{i=a,b}\sum_{\kv} \frac{1}{3!}(\partial_x\ep_i\partial_y\ep_i)^2\partial_{\ep_i}^3 f_i
    \nonumber\\
    &\ra
    \frac{1}{2}
    \frac{1}{3!}
    \frac{e^2}{\hbar^2 c^2}\sum_{i=a,b}\sum_\kv
    \frac{\partial}{\partial k_{y}}\left[
    \lb\frac{\partial\ep_i}{\partial k_{y}}\rb^{-1} 
    \frac{\partial}{\partial k_{x}}\left[(\partial_x\ep_i\partial_y\ep_i)^2
    \lb\frac{\partial\ep_i}{\partial k_{x}}\rb^{-1} 
    \right] 
    \right]
    f'_i
    +(x\leftrightarrow y)
    \nonumber\\
    &=
    \frac{1}{2}\frac{1}{3!}
    \frac{e^2}{\hbar^2 c^2}\sum_{i=a,b}\sum_\kv
    \left[
    2\lb
    \frac{\partial^2\ep_i}
    {\partial k_x\partial k_y}
    \rb^2
    +
    2\frac{\partial\ep_i}
    {\partial k_x}
    \frac{\partial^3\ep_i}
    {\partial k_x\partial k_y^2}
    +
    \frac{\partial\ep_i}
    {\partial k_y}
    \frac{\partial^3\ep_i}
    {\partial k_x^2\partial k_y}
    +
    \frac{\partial^2\ep_i}
    {\partial k_x^2}
    \frac{\partial^2\ep_i}
    {\partial k_y^2}
    \right]
    f'_i
    +(x\leftrightarrow y)
    \nonumber\\
    &=
    \frac{1}{3!}
    \frac{e^2}{\hbar^2 c^2}\sum_{i=a,b}\sum_\kv
    \left[
    2\lb
    \frac{\partial^2\ep_i}
    {\partial k_x\partial k_y}
    \rb^2
    +
    \frac{\partial^2\ep_i}
    {\partial k_x^2}
    \frac{\partial^2\ep_i}
    {\partial k_y^2}
    +
    \frac{3}{2}
    \lb\frac{\partial\ep_i}
    {\partial k_x}
    \frac{\partial^3\ep_i}
    {\partial k_x\partial k_y^2}
    +
    \frac{\partial\ep_i}
    {\partial k_y}
    \frac{\partial^3\ep_i}
    {\partial k_x^2\partial k_y}\rb
    \right]
    f'_i\label{chiE},
\end{align}
which was previously derived in Ref.~\cite{Fukuyama1971}. We then note that $\chi_\geo^{(1)}$ is vanishing, and $\chi_\geo$ can be rewritten as:
\begin{align}
    \chi_{xy} &=
    4\frac{e^2}{\hbar^2 c^2}k_BT\sum_n\sum_\kv \frac{\ep_{ab}^2\partial_x\ep_b\partial_y\ep_b}
    {(i\om_n+\mu-\ep_b)^3(i\om_n+\mu-\ep_a)}
    g_{xy}^{ab} + (a\leftrightarrow b)
    \nonumber\\
    &=
    4\frac{e^2}{\hbar^2 c^2}\sum_\kv \ep_{ab}^2\partial_x\ep_a\partial_y\ep_a
    g_{xy}^{ab}
    \lb
    \frac{f_{ab}}{\ep_{ab}^3}
    -
    \frac{f_{a}'}{\ep_{ab}^2}
    +
    \frac{f_{a}''}{2\ep_{ab}}
    \rb
    + (a\leftrightarrow b)
    \nonumber\\
    &=
    4\frac{e^2}{\hbar^2 c^2}\sum_\kv \frac{1}{\ep_{ab}}
    \partial_x\ep_a\partial_y\ep_a
    g_{xy}^{ab}
    \lb
    f_{ab}
    -
    \ep_{ab}f_{a}'
    +
    \frac{1}{2}\ep_{ab}^2f_{a}''
    \rb
    + (a\leftrightarrow b)
    \ ,
    \end{align}
    \begin{align}
    \chi_{xx}
    &=
    2\frac{e^2}{\hbar^2 c^2}k_BT\sum_n\sum_\kv
    \frac{\ep_{ab}^2\partial_y\ep_a\partial_y\ep_b}
    {(i\om_n+\mu-\ep_a)^2(i\om_n+\mu-\ep_b)^2} g_{xx}^{ab} 
    \nonumber\\
    &=
    2\frac{e^2}{\hbar^2 c^2}\sum_\kv
    {\ep_{ab}^2\partial_y\ep_a\partial_y\ep_b} g_{xx}^{ab} 
    \frac{1}{\ep_{ab}^{2}}\lb
    f'_a+f'_b
    -
    2\frac{f_{ab}}{\ep_{ab}}
    \rb\
    \nonumber\\
    &=
    2\frac{e^2}{\hbar^2 c^2}\sum_\kv
    \partial_y\ep_a\partial_y\ep_b g_{xx}^{ab} 
    \lb
    f'_a+f'_b
    -
    2\frac{f_{ab}}{\ep_{ab}}
    \rb\ ,
    \\
    \chi_{yy}
    &=
    2\frac{e^2}{\hbar^2 c^2}\sum_\kv
    \partial_x\ep_a\partial_x\ep_b g_{yy}^{ab} 
    \lb
    f'_a+f'_b
    -
    2\frac{f_{ab}}{\ep_{ab}}
    \rb\ .
    \label{chigeo}
\end{align}
\subsubsection{Beyond two-band limit}
Beyond two-band terms, we note that Eqs.~(\ref{chiE}-\ref{chigeo}) can be generalized to any number of bands naturally, by performing $\frac{1}{2}\sum_{a\neq b}$, without any change in the function form. However, $\chi_\geo^{(1)}$ is now nonvanishing:
\begin{align}
    \chi_\geo^{(1)}
    &=
    4\frac{e^2}{\hbar^2c^2}k_BT\sum_n\sum_\kv\sum_{a,b,c}
    \frac{\ep_{ba}\ep_{ac}\partial_y\ep_a}
    {(i\om_n+\mu-\ep_a)^2(i\om_n+\mu-\ep_c)}
    \Re\Big[\braket{a}{\partial_x b}\braket{b}{\partial_y c}\braket{c}{\partial_x a}\Big]
    + (x\leftrightarrow y)
    \nonumber\\
    &=
    4\frac{e^2}{\hbar^2c^2}\sum_\kv\sum_{a,b,c}
    {\ep_{ba}\ep_{ac}\partial_y\ep_a}
    \frac{1}{\ep_{ac}}
    \lb
    f'_a - \frac{f_{ac}}{\ep_{ac}}
    \rb
    \Re\Big[\braket{a}{\partial_x b}\braket{b}{\partial_y c}\braket{c}{\partial_x a}\Big]
    + (x\leftrightarrow y)
    \nonumber\\
    &=
    -
    4\frac{e^2}{\hbar^2c^2}\sum_\kv\sum_{a,b,c}
    {\ep_{ca}\ep_{ab}\partial_y\ep_a}
    \frac{1}{\ep_{ab}}
    \lb
    f'_a - \frac{f_{ab}}{\ep_{ab}}
    \rb
    \Re\Big[\braket{a}{\partial_x b}\braket{b}{\partial_y c}\braket{c}{\partial_x a}\Big]
    + (x\leftrightarrow y)
    \nonumber\\
    &=
    2\frac{e^2}{\hbar^2c^2}\sum_\kv\sum_{a,b,c}
    {\partial_y\ep_a}
    \lb
    \ep_{bc}f'_a + 
    \frac{\ep_{ca}^2f_{ab}+\ep_{ab}^2f_{ca}}
    {\ep_{ab}\ep_{ca}}
    \rb
    \Re\Big[\braket{a}{\partial_x b}\braket{b}{\partial_y c}\braket{c}{\partial_x a}\Big]
    + (x\leftrightarrow y)
    \nonumber\\
    &=
    2\frac{e^2}{\hbar^2c^2}\sum_\kv\sum_{a,b,c}
    {\partial_y\ep_a}
    \lb
    \ep_{bc}f'_a + 
    \ep_{ca}\frac{f_{ab}}{\ep_{ab}}
    +
    \ep_{ab}\frac{f_{ca}}{\ep_{ca}}
    \rb
    \Re\Big[\braket{a}{\partial_x b}\braket{b}{\partial_y c}\braket{c}{\partial_x a}\Big]
    + (x\leftrightarrow y)
    \nonumber\\
    &=
    2\frac{e^2}{\hbar^2c^2}\sum_\kv\sum_{a,b,c}
    {\partial_y\ep_a}
    \lb
    \ep_{bc}f'_a + 
    2\ep_{ca}\frac{f_{ab}}{\ep_{ab}}
    \rb
    \Re\Big[\braket{a}{\partial_x b}\braket{b}{\partial_y c}\braket{c}{\partial_x a}\Big]
    + (x\leftrightarrow y)
    \ ,
\end{align}
and also $\chi_\geo^{(2)}$ has the possibility of involving two, three or four bands, which should be considered separately. The two-band case is identical to Eq.~\eqref{chigeo2_2}, where indices $(a,\ c)$ and $(b,\ d)$ are matched pairwise. For the three-band case, we match $(a,\ c)$ pairwise (matching $(b,\ d)$ pairwise gives the same result by symmetry, contributing a factor of two), which yields:
\begin{align}
    \chi^{(2)}_\geo
    &=
    2\frac{e^2}{\hbar^2c^2}k_BT\sum_n\sum_\kv\sum_{a,b,d} \Re\Big[
    \braket{a}{\partial_y b}\braket{b}{\partial_x a}\braket{a}{\partial_y d}\braket{d}{\partial_x a}\Big]
    \frac{\ep_{ab}\ep_{ba}\ep_{ad}\ep_{da}}
    {(i\om_n+\mu-\ep_a)^2(i\om_n+\mu-\ep_b)(i\om_n+\mu-\ep_d)}
    \nonumber\\
    &=
    2\frac{e^2}{\hbar^2c^2}\sum_\kv\sum_{a,b,d} 
    \lb
    g_{xy}^{ab}g_{xy}^{ad}
    -
    \frac{\Om_{xy}^{ab}\Om_{xy}^{ad}}{4}
    \rb
    \ep_{ab}^2\ep_{ad}^2
    \lb
    \frac{f'_a}{\ep_{ab}\ep_{ad}}
    -
    \frac{f_a}{\ep_{ab}^2\ep_{ad}}
    -
    \frac{f_a}{\ep_{ab}\ep_{ad}^2}
    +
    \frac{f_b}{\ep_{ba}^2\ep_{bd}}
    +
    \frac{f_d}{\ep_{da}^2\ep_{db}}
    \rb
    \nonumber\\
    &=
    2\frac{e^2}{\hbar^2c^2}\sum_\kv\sum_{a,b,d} 
    \lb
    g_{xy}^{ab}g_{xy}^{ad}
    -
    \frac{\Om_{xy}^{ab}\Om_{xy}^{ad}}{4}
    \rb
    \lb
    {f'_a}\ep_{ab}\ep_{ad}
    -
    \frac{\ep_{ab}^2f_{da}+\ep_{da}^2f_{ab}}{\ep_{bd}}
    \rb
    \nonumber\\
    &=
    2\frac{e^2}{\hbar^2c^2}\sum_\kv\sum_{a,b,d} 
    \lb
    g_{xy}^{ab}g_{xy}^{ad}
    -
    \frac{\Om_{xy}^{ab}\Om_{xy}^{ad}}{4}
    \rb
    \lb
    {f'_a}\ep_{ab}\ep_{ad}
    -
    \frac{2\ep_{ab}^2f_{da}}{\ep_{bd}}
    \rb,
\end{align}
and if four-band contributions are involved, we instead have:
\begin{align}
    \chi_\geo^{(2)}
    &=
    \frac{e^2}{\hbar^2c^2}\sum_\kv\sum_{a,b,c,d} \Re\Big[
    \braket{a}{\partial_y b}\braket{b}{\partial_x c}\braket{c}{\partial_y d}\braket{d}{\partial_x a}\Big]
    \lb
    -
    \ep_{cd}\ep_{bc}\frac{f_a}{\ep_{ac}}
    -
    \ep_{da}\ep_{cd}\frac{f_b}{\ep_{bd}}
    +
    \ep_{ab}\ep_{da}\frac{f_c}{\ep_{ac}}
    +
    \ep_{ab}\ep_{bc}\frac{f_d}{\ep_{bd}}
    \rb
    \nonumber\\
    &=
    -4\frac{e^2}{\hbar^2c^2}\sum_\kv\sum_{a,b,c,d} \Re\Big[
    \braket{a}{\partial_y b}\braket{b}{\partial_x c}\braket{c}{\partial_y d}\braket{d}{\partial_x a}\Big]
    \ep_{cd}\ep_{bc}\frac{f_a}{\ep_{ac}}
    \ .
\end{align}
\subsection{Beyond the Fukuyama formula}
    In Refs.~\cite{Raoux2015, Gomez2011}, it was shown that there is an extra contribution $\tilde\chi_\mathrm{O}$ to the orbital magnetic susceptibility, beyond the Fukuyama formula Eq.~\eqref{Fukuyama}, when the Hamiltonian is not of the canonical form,
    \beq{}
        \tilde\chi_\mathrm{O}=
        \frac{e^2}{\hbar^2 c^2}k_BT\sum_n\sum_\kv
        \frac{1}{2} 
        \Tr[(Gj_xGj_y+Gj_yGj_x)Gj_{xy}]
    \eep
    Using Eq.~\eqref{GjG}, we have
    \begin{align}
        \Tr[(Gj_xGj_y+Gj_yGj_x)Gj_{xy}]
        &=\Tr[(\partial_xGj_y+\partial_yGj_x)Gj_{xy}]
        \nonumber\\
        &=\Tr[\partial_xGj_yGj_{xy}]
        +(x\leftrightarrow y).
    \end{align}
    In band basis, we note that $j_{xy}=\partial_x\partial_y H$ can be written as
    \begin{align}
        j_{xy}
        &=
        \sum_a
        \partial_x(\partial_y \ep_a P_a
        +\ep_a \partial_y P_a)
        \nonumber\\
        &=
        \sum_a
        \partial_x \partial_y \ep_a P_a
        +\partial_y \ep_a \partial_x P_a
        +\partial_x \ep_a \partial_y P_a
        +\ep_a \partial_x\partial_y P_a\;.
    \end{align}
    As such, we can expand the extra term $\tilde\chi_\mathrm{O}$ correspondingly as
    \begin{align}
         \tilde\chi_\mathrm{O}
         &=
         \frac{e^2}{2\hbar^2 c^2}k_BT\sum_n\sum_\kv
         \mathrm{Tr}[\partial_xGj_yGj_{xy}]
         + (x\leftrightarrow y)
         \nonumber\\
         &=
         \frac{e^2}{2\hbar^2 c^2}k_BT
         \sum_n\sum_\kv\sum_{a,b,c,d}
         \Tr\bigg[
         \left(\frac{P_a\partial_x\ep_a}{(i\om_n+\mu-\ep_a)^2}
         +\frac{\partial_xP_a}{i\om_n+\mu-\ep_a}\right)
         (\partial_y\ep_bP_b+\ep_b\partial_yP_b)
         \frac{P_c}{i\om_n+\mu-\ep_c}
         \Big]
         \nonumber\\
         &\phantom{\frac{}{}}
         (\partial_x \partial_y \ep_d P_d
         +\partial_y \ep_d \partial_x P_d
         +\partial_x \ep_d \partial_y P_d
         +\ep_d \partial_x\partial_y P_d)
         +(x\leftrightarrow y)\bigg]
         \nonumber\\
         &=
         \frac{e^2}{2\hbar^2 c^2}k_BT
         \sum_n\sum_\kv\sum_{a,b,c,d} 
         \frac{\partial_x\ep_a\partial_y\ep_b\partial_x\partial_y\ep_d}
         {(i\om_n+\mu-\ep_a)^2(i\om_n+\mu-\ep_c)}
         \Tr[P_aP_bP_cP_d]
         +
         \frac{\ep_d\partial_x\ep_a\partial_y\ep_b}
         {(i\om_n+\mu-\ep_a)^2(i\om_n+\mu-\ep_c)}
         \Tr[P_aP_bP_c\partial_x\partial_yP_d]
         \nonumber\\
         &
         +\frac{\ep_b\partial_x\partial_y\ep_d}
         {(i\om_n+\mu-\ep_a)(i\om_n+\mu-\ep_c)}
         \Tr[\partial_xP_a\partial_yP_bP_cP_d]
         +\frac{\partial_y\ep_b\partial_x\ep_d}
         {(i\om_n+\mu-\ep_a)(i\om_n+\mu-\ep_c)}
         \Tr[\partial_xP_aP_bP_c\partial_yP_d]
         \nonumber\\
         &
         +\frac{\ep_b\partial_x\ep_a\partial_y\ep_d}
         {(i\om_n+\mu-\ep_a)^2(i\om_n+\mu-\ep_c)}
         \Tr[P_a\partial_yP_bP_c\partial_xP_d]
         +\frac{\ep_b\partial_x\ep_a\partial_x\ep_d}
         {(i\om_n+\mu-\ep_a)^2(i\om_n+\mu-\ep_c)}
         \Tr[P_a\partial_yP_bP_c\partial_yP_d]
         \nonumber\\
         &+\frac{\partial_y\ep_b\partial_y\ep_d}
         {(i\om_n+\mu-\ep_a)(i\om_n+\mu-\ep_c)}
         \Tr[\partial_xP_aP_bP_c\partial_xP_d]
         +\frac{\ep_d\partial_y\ep_b}
         {(i\om_n+\mu-\ep_a)(i\om_n+\mu-\ep_c)}
         \Tr[\partial_xP_aP_bP_c\partial_x\partial_yP_d]
         \nonumber\\
         &+\frac{\ep_b\ep_d\partial_x\ep_a}
         {(i\om_n+\mu-\ep_a)^2(i\om_n+\mu-\ep_c)}
         \Tr[P_a\partial_yP_bP_c\partial_x\partial_yP_d]
         +\frac{\ep_b\partial_y\ep_d}
         {(i\om_n+\mu-\ep_a)(i\om_n+\mu-\ep_c)}
         \Tr[\partial_xP_a\partial_yP_bP_c\partial_xP_d]
         \nonumber\\
         &+\frac{\ep_b\partial_x\ep_d}
         {(i\om_n+\mu-\ep_a)(i\om_n+\mu-\ep_c)}
         \Tr[\partial_xP_a\partial_yP_bP_c\partial_yP_d]
         +\frac{\ep_b\ep_d}
         {(i\om_n+\mu-\ep_a)(i\om_n+\mu-\ep_c)}
         \Tr[\partial_xP_a\partial_yP_bP_c\partial_x\partial_yP_d]
         +(x\leftrightarrow y)\;,
    \end{align}
    where we have used Eq.~\eqref{Tr1} to remove vanishing terms. Below, we will refer to each term as $\tilde\chi_1,\tilde\chi_2,\cdots\tilde\chi_{12}$ . To further simplify, we first employ additional sum rules of projector matrices and their derivatives:
    \begin{align}
        \sum_{a,b,c,d}\Tr[P_aP_bP_c\partial_x\partial_yP_d]
        &=
        \sum_{a,b,c,d}\d_{ab}\d_{ac}\Tr[P_a\partial_x\partial_yP_d]
        \nonumber\\
        &=
        \sum_{a,b,c,d}\d_{ab}\d_{ac}\lb\braket{d}{\partial_x\partial_y d}\d_{ad}
        +\braket{a}{\partial_y d}\braket{\partial_x d}{a}
        +\braket{a}{\partial_x d}\braket{\partial_y d}{a}
        +\braket{\partial_x\partial_y d}{d}\d_{ad}
        \rb
        \nonumber\\
        &=
        \sum_{a,b,c,d}\d_{ab}\d_{ac}\left[
        \braket{\partial_y a}{d}\braket{d}{\partial_x a}
        +\braket{\partial_x a}{d}\braket{d}{\partial_y a}
        -\lb\braket{\partial_x a}{\partial_y a}
        +\braket{\partial_y a}{\partial_x a}\rb\d_{ad}
        \right]\;,
        \label{Pxy}
    \end{align}
    \begin{align}
        \sum_{a,b,c,d}\Tr[P_a\partial_\mu P_bP_c\partial_\nu P_d]
        &=
        \sum_{a,b,c,d}\Tr[P_a\partial_\mu P_b\partial_\nu(P_cP_d)
        -P_a\partial_\mu P_b\partial_\nu P_c P_d]
        \nonumber\\
        &=
        \sum_{a,b,c,d}\Tr[P_a\partial_\mu P_b\partial_\nu P_c\d_{cd}
        -P_a\partial_\mu P_b\partial_\nu P_c \d_{ad}]
        \nonumber\\
        &=
        \sum_{a,b,c,d}\Tr[P_a\partial_\mu P_b\partial_\nu P_c
        (\d_{cd}-\d_{ad})]
        \nonumber\\
        &=
        \sum_{a,b,c,d}\lb-\xi^{ab}_\mu\xi^{ba}_\nu\d_{ac}
        +\xi^{ab}_\mu\xi^{ba}_\nu\d_{bc}
        +\braket{\partial_\mu a}{\partial_\nu a}\d_{ab}\d_{ac}
        -\xi^{ac}_\mu\xi^{ca}_\nu\d_{ab}\rb
        (\d_{cd}-\d_{ad})
        \nonumber\\
        &=
        \sum_{a,b,c,d}\xi^{ac}_\mu\xi^{ca}_\nu
        (\d_{bc}-\d_{ab})(\d_{cd}-\d_{ad}).
        \label{PmuPnu_2}
    \end{align}
    We begin by calculating $\tilde\chi_1$, which is a single band contribution, thus a correction term to the energetic term amounts to:
    \begin{align}
        \tilde\chi_1 &= 
        \frac{e^2}{2\hbar^2 c^2}k_BT
        \sum_n\sum_\kv\sum_{a,b,c,d} 
        \frac{\partial_x\ep_a\partial_y\ep_b\partial_x\partial_y\ep_d}
        {(i\om_n+\mu-\ep_a)^2(i\om_n+\mu-\ep_c)}
        \Tr[P_aP_bP_cP_d]
        + (x\leftrightarrow y)
        \nonumber\\
        &=
        \frac{e^2}{\hbar^2 c^2}k_BT
        \sum_n\sum_\kv\sum_{a} 
        \frac{\partial_x\ep_a\partial_y\ep_a\partial_x\partial_y\ep_a}
        {(i\om_n+\mu-\ep_a)^3}
        \nonumber\\
        &=
        \frac{1}{2!}
        \frac{e^2}{\hbar^2 c^2}
        \sum_\kv\sum_{a} 
        f''_a\partial_x\ep_a\partial_y\ep_a\partial_x\partial_y\ep_a
        \nonumber\\
        &=
        -\frac{1}{2}\frac{e^2}{2\hbar^2 c^2}
        \sum_\kv\sum_{a} 
        f'_a\partial_x(\partial_y\ep_a\partial_x\partial_y\ep_a)
        + (x\leftrightarrow y)
        \nonumber\\
        &=
        -\frac{e^2}{4\hbar^2 c^2}
        \sum_\kv\sum_{a} 
        \left[
        2\lb
        \frac{\partial^2\ep_a}
        {\partial k_x\partial k_y}
        \rb^2
        +
        \lb\frac{\partial\ep_a}
        {\partial k_x}
        \frac{\partial^3\ep_a}
        {\partial k_x\partial k_y^2}
        +
        \frac{\partial\ep_a}
        {\partial k_y}
        \frac{\partial^3\ep_a}
        {\partial k_x^2\partial k_y}\rb
        \right]
        f'_a\; ,
    \end{align}
    where we have used Eq.~\eqref{MS} and Eq.~\eqref{fpp} to simplify the expression. Combining with $\chi_\text{E}$, Eq.~\eqref{chiE}, the total energetic contribution is given by the Landau-Peierls formula \cite{Peierls1933}:
    \beq{LP}
        \chi_\mathrm{LP}
        =
        \frac{e^2}{6\hbar^2c^2}
        \sum_\kv\sum_a
        \left[
        \frac{\partial^2\ep_a}
        {\partial k_x^2}
        \frac{\partial^2\ep_a}
        {\partial k_y^2}
        -
        \lb
        \frac{\partial^2\ep_a}
        {\partial k_x\partial k_y}
        \rb^2
        \right]
        f'_a
    \eep
    Next, we calculate $\tilde\chi_2,\cdots \tilde\chi_5$, which involve only two bands and are proportional to $g^{ab}_{xy}$. $\tilde{\chi}_2$ and $\tilde{\chi}_3$ can be simplified as
    \begin{align}
        \tilde\chi_2
        &=
        \frac{e^2}{2\hbar^2 c^2}k_BT
        \sum_n\sum_\kv\sum_{a,b,c,d}
        \frac{\ep_d\partial_x\ep_a\partial_y\ep_b}
        {(i\om_n+\mu-\ep_a)^2(i\om_n+\mu-\ep_c)}
        \Tr[P_aP_bP_c\partial_x\partial_yP_d]
        + (x \leftrightarrow y)
        \nonumber\\
        &=
        \frac{e^2}{2\hbar^2 c^2}k_BT
        \sum_n\sum_\kv\sum_{a,d}
        \frac{\ep_d\partial_x\ep_a\partial_y\ep_a}
        {(i\om_n+\mu-\ep_a)^3}
        \left[
        \braket{\partial_y a}{d}\braket{d}{\partial_x a}
        +\braket{\partial_x a}{d}\braket{d}{\partial_y a}
        -\lb\braket{\partial_x a}{\partial_y a}
        +\braket{\partial_y a}{\partial_x a}\rb\d_{ad}
        \right]
        + (x\leftrightarrow y)
        \nonumber\\
        &=
        -\frac{e^2}{\hbar^2 c^2}k_BT
        \sum_n\sum_\kv\sum_{a,b}
        \frac{\partial_x\ep_a\partial_y\ep_a}
        {(i\om_n+\mu-\ep_a)^3}
        \lb
        \braket{\partial_y a}{b}\braket{b}{\partial_x a}
        +\braket{\partial_x a}{b}\braket{b}{\partial_y a}
        \rb
        \ep_{ab}
        \nonumber\\
        &=
        -2\frac{e^2}{\hbar^2 c^2}k_BT
        \sum_n\sum_\kv\sum_{a,b}
        \frac{\ep_{ab}\partial_x\ep_a\partial_y\ep_a}
        {(i\om_n+\mu-\ep_a)^3}
        g_{xy}^{ab}
        \;,\label{tchi2}
    \end{align}
    \begin{align}
        \tilde\chi_3
        &=
        \frac{e^2}{2\hbar^2 c^2}k_BT
        \sum_n\sum_\kv\sum_{a,b,c,d}
        \frac{\ep_b\partial_x\partial_y\ep_d}
        {(i\om_n+\mu-\ep_a)(i\om_n+\mu-\ep_c)}
        \Tr[\partial_xP_a\partial_yP_bP_cP_d]
        + (x \leftrightarrow y)
        \nonumber\\
        &=
        \frac{e^2}{2\hbar^2 c^2}k_BT
        \sum_n\sum_\kv\sum_{a,b,c}
        \frac{\ep_b\partial_x\partial_y\ep_c}
        {(i\om_n+\mu-\ep_a)(i\om_n+\mu-\ep_c)}
        \lb-\xi_x^{ba}\xi_y^{ab}\d_{bc}
        +\xi_x^{ca}\xi_y^{ac}\d_{ab}
        +\braket{\partial_x a}{\partial_y a}\d_{bc}\d_{ac}
        -\xi_x^{ab}\xi_y^{ba}\d_{ac}
        \rb
        + (x \leftrightarrow y)
        \nonumber\\
        &=
        \frac{e^2}{2\hbar^2 c^2}k_BT
        \sum_n\sum_\kv\sum_{a,b}
        2\Re[\xi_x^{ba}\xi_y^{ab}]
        \lb
        \frac{\ep_{ab}\partial_x\partial_y\ep_b}
        {(i\om_n+\mu-\ep_a)(i\om_n+\mu-\ep_b)}
        +
        \frac{\ep_{ab}\partial_x\partial_y\ep_a}
        {(i\om_n+\mu-\ep_a)^2}
        \rb
        \nonumber\\
        &=
        \frac{e^2}{\hbar^2 c^2}k_BT
        \sum_n\sum_\kv\sum_{a,b}
        g_{xy}^{ab}
        \ep_{ab}\partial_x\partial_y\ep_b
        \lb
        \frac{1}
        {(i\om_n+\mu-\ep_a)(i\om_n+\mu-\ep_b)}
        -
        \frac{1}
        {(i\om_n+\mu-\ep_b)^2}
        \rb
        \nonumber\\
        &=
        \frac{e^2}{\hbar^2 c^2}k_BT
        \sum_n\sum_\kv\sum_{a,b}
        \frac{\ep_{ab}^2\partial_x\partial_y\ep_b}
        {(i\om_n+\mu-\ep_a)(i\om_n+\mu-\ep_b)^2}
        g^{ab}_{xy}
        \;.
        \label{tchi4}
    \end{align}
    To simplify $\tilde\chi_4$ and $\tilde\chi_5$, we combine both terms together. To be specific, $\tilde\chi_4$ and $\tilde\chi_5$ are the complex conjugates of each other:
    \begin{align}
        \tilde\chi_4
        &=
        \frac{e^2}{2\hbar^2 c^2}k_BT
        \sum_n\sum_\kv\sum_{a,b,c,d}
        \frac{\partial_y\ep_b\partial_x\ep_d}
        {(i\om_n+\mu-\ep_a)(i\om_n+\mu-\ep_c)}
        \Tr[\partial_xP_aP_bP_c\partial_yP_d]
        + (x \leftrightarrow y)
        \nonumber\\
        &=
        \frac{e^2}{2\hbar^2 c^2}k_BT
        \sum_n\sum_\kv\sum_{a,b,c,d}
        \frac{\partial_x\ep_c\partial_y\ep_a}
        {(i\om_n+\mu-\ep_b)(i\om_n+\mu-\ep_d)}
        \Tr[\partial_xP_a\partial_yP_bP_cP_d]
        + (x \leftrightarrow y)
        \nonumber\\
        &=
        \frac{e^2}{2\hbar^2 c^2}k_BT
        \sum_n\sum_\kv\sum_{a,b,c}
        \frac{\partial_x\ep_c\partial_y\ep_a}
        {(i\om_n+\mu-\ep_b)(i\om_n+\mu-\ep_c)}
        \lb-\xi_x^{ba}\xi_y^{ab}\d_{bc}
        +\xi_x^{ca}\xi_y^{ac}\d_{ab}
        +\braket{\partial_x a}{\partial_y a}\d_{bc}\d_{ac}
        -\xi_x^{ab}\xi_y^{ba}\d_{ac}
        \rb
        + (x \leftrightarrow y)
        \nonumber\\
        &=
        \frac{e^2}{2\hbar^2 c^2}k_BT
        \sum_n\sum_\kv\sum_{a,b}
        \xi^{ab}_y\xi^{ba}_x
        \left[
        -\frac{\partial_x\ep_b\partial_y\ep_a}
        {(i\om_n+\mu-\ep_b)^2}
        +\frac{\partial_y\ep_a(\partial_x\ep_b-\partial_x\ep_a)}
        {(i\om_n+\mu-\ep_a)(i\om_n+\mu-\ep_b)}
        +\frac{\partial_x\ep_a\partial_y\ep_a}
        {(i\om_n+\mu-\ep_a)^2}
        \right]
        +(x\leftrightarrow y)
        \nonumber\\
        &=
        \frac{e^2}{2\hbar^2 c^2}k_BT
        \sum_n\sum_\kv\sum_{a,b}
        \xi^{ab}_y\xi^{ba}_x
        \left[
        \frac{\partial_y\ep_a(\partial_x\ep_a-\partial_x\ep_b)}
        {(i\om_n+\mu-\ep_a)^2}
        +\frac{\partial_y\ep_a(\partial_x\ep_b-\partial_x\ep_a)}
        {(i\om_n+\mu-\ep_a)(i\om_n+\mu-\ep_b)}
        \right]
        +(x\leftrightarrow y)
        \;,\\
        \tilde\chi_5
        &=
        \frac{e^2}{2\hbar^2 c^2}k_BT
        \sum_n\sum_\kv\sum_{a,b,c,d}
        \frac{\ep_b\partial_x\ep_a\partial_y\ep_d}
        {(i\om_n+\mu-\ep_a)^2(i\om_n+\mu-\ep_c)}
        \Tr[P_a\partial_yP_bP_c\partial_xP_d]
        + (x \leftrightarrow y)
        \nonumber\\
        &=
        \frac{e^2}{2\hbar^2 c^2}k_BT
        \sum_n\sum_\kv\sum_{a,b,c,d}
        \frac{\ep_b\partial_x\ep_a\partial_y\ep_d}
        {(i\om_n+\mu-\ep_a)^2(i\om_n+\mu-\ep_c)}
        \xi^{ac}_y\xi^{ca}_x
        (\d_{bc}-\d_{ab})(\d_{cd}-\d_{ad})
        +(x\leftrightarrow y)
        \nonumber\\
        &=
        \frac{e^2}{2\hbar^2 c^2}k_BT
        \sum_n\sum_\kv\sum_{a,b}
        \xi^{ab}_y\xi^{ba}_x
        \frac{\ep_{ab}\partial_x\ep_a(\partial_y\ep_a-\partial_y\ep_b)}
        {(i\om_n+\mu-\ep_a)^2(i\om_n+\mu-\ep_b)}
        +(x\leftrightarrow y)
        \nonumber\\
        &=
        \frac{e^2}{2\hbar^2 c^2}k_BT
        \sum_n\sum_\kv\sum_{a,b}
        \xi^{ab}_y\xi^{ba}_x
        \left[
        \frac{\partial_x\ep_a(\partial_y\ep_a-\partial_y\ep_b)}
        {(i\om_n+\mu-\ep_a)^2}
        +
        \frac{\partial_x\ep_a(\partial_y\ep_b-\partial_y\ep_a)}
        {(i\om_n+\mu-\ep_a)(i\om_n+\mu-\ep_b)}
        \right]
        +(x\leftrightarrow y)
        \nonumber\\
        &=
        \tilde\chi_4^*\;.
    \end{align}
    As such, combining $\tilde{\chi}_4$ and $\tilde\chi_5$ gives:
    \begin{align}
        \tilde\chi_4 + \tilde\chi_5
        &=
        \frac{e^2}{2\hbar^2 c^2}k_BT
        \sum_n\sum_\kv\sum_{a,b}
        (\xi^{ab}_y\xi^{ba}_x+\xi^{ab}_x\xi^{ba}_y)
        \left[
        \frac{\partial_x\ep_a(\partial_y\ep_a-\partial_y\ep_b)}
        {(i\om_n+\mu-\ep_a)^2}
        +
        \frac{\partial_x\ep_a(\partial_y\ep_b-\partial_y\ep_a)}
        {(i\om_n+\mu-\ep_a)(i\om_n+\mu-\ep_b)}
        +(x\leftrightarrow y)
        \right]
        \nonumber\\
        &=
        \frac{e^2}{\hbar^2 c^2}k_BT
        \sum_n\sum_\kv\sum_{a,b}
        g_{xy}^{ab}
        \frac{2\partial_x\ep_a\partial_y\ep_a
        -\partial_x\ep_a\partial_y\ep_b
        -\partial_x\ep_b\partial_y\ep_a}
        {i\om_n-\mu-\ep_a}
        \lb
        \frac{1}{i\om_n-\mu-\ep_a}
        -\frac{1}{i\om_n-\mu-\ep_b}
        \rb
        \nonumber\\
        &=
        \frac{e^2}{\hbar^2 c^2}k_BT
        \sum_n\sum_\kv\sum_{a,b}
        g_{xy}^{ab}
        \frac{\ep_{ab}
        (2\partial_x\ep_a\partial_y\ep_a
        -\partial_x\ep_a\partial_y\ep_b
        -\partial_x\ep_b\partial_y\ep_a)}
        {(i\om_n-\mu-\ep_a)^2(i\om_n-\mu-\ep_b)}
        \nonumber\\
        &=
        \frac{e^2}{\hbar^2 c^2}k_BT
        \sum_n\sum_\kv\sum_{a,b}
        \left[
        \frac{2\ep_{ab}\partial_x\ep_a\partial_y\ep_a}
        {(i\om_n-\mu-\ep_a)^2(i\om_n-\mu-\ep_b)}
        -
        \frac{\ep_{ab}^2(\partial_x\ep_a\partial_y\ep_b
        +\partial_x\ep_b\partial_y\ep_a)}
        {2(i\om_n-\mu-\ep_a)^2(i\om_n-\mu-\ep_b)^2}
        \right]
        g_{xy}^{ab}\;.
    \end{align}
    Overall, additional contribution $\tilde\chi_{xy}$ proportional to $g_{xy}^{ab}$ is given by
    \begin{align}
        \tilde\chi_{xy} &=
        \tilde\chi_2 + \tilde\chi_3 + \tilde\chi_4 + \tilde\chi_5 
        \nonumber\\
        &=
        \frac{e^2}{\hbar^2 c^2}k_BT
        \sum_n\sum_\kv\sum_{a,b}
        \left[
        \frac{2\ep_{ab}\partial_x\ep_a\partial_y\ep_a}
        {(i\om_n-\mu-\ep_a)^2}
        \lb
        \frac{1}{i\om_n-\mu-\ep_b}
        -\frac{1}{i\om_n-\mu-\ep_a}
        \rb
        +
        \frac{\ep_{ab}^2\partial_x\partial_y\ep_b}
        {(i\om_n+\mu-\ep_a)(i\om_n+\mu-\ep_b)^2}
        \right.
        \nonumber\\
        &-\left.
        \frac{\ep_{ab}^2(\partial_x\ep_a\partial_y\ep_b
        +\partial_x\ep_b\partial_y\ep_a)}
        {2(i\om_n-\mu-\ep_a)^2(i\om_n-\mu-\ep_b)^2}
        \right]
        g_{xy}^{ab}
        \nonumber\\
        &=
        \frac{e^2}{\hbar^2 c^2}k_BT
        \sum_n\sum_\kv\sum_{a,b}
        \left[
        -\frac{2\ep_{ab}^2\partial_x\ep_a\partial_y\ep_a}
        {(i\om_n-\mu-\ep_a)^3(i\om_n-\mu-\ep_b)}
        +
        \frac{\ep_{ab}^2\partial_x\partial_y\ep_a}
        {(i\om_n+\mu-\ep_a)^2(i\om_n+\mu-\ep_b)}
        -
        \frac{\ep_{ab}^2(\partial_x\ep_a\partial_y\ep_b
        +\partial_x\ep_b\partial_y\ep_a)}
        {2(i\om_n-\mu-\ep_a)^2(i\om_n-\mu-\ep_b)^2}
        \right]
        g_{xy}^{ab}\;.
    \end{align}
    Combining with $\chi_{xy}$, we have
    \begin{align}
        &\chi_{xy} + \tilde\chi_{xy}
        \nonumber\\
        &=
        \frac{e^2}{\hbar^2 c^2}k_BT
        \sum_n\sum_\kv\sum_{a,b}
        \left[
        \frac{2\ep_{ab}^2\partial_x\ep_a\partial_y\ep_a}
        {(i\om_n-\mu-\ep_a)^3(i\om_n-\mu-\ep_b)}
        +
        \frac{\ep_{ab}^2\partial_x\partial_y\ep_a}
        {(i\om_n+\mu-\ep_a)^2(i\om_n+\mu-\ep_b)}
        -
        \frac{\ep_{ab}^2(\partial_x\ep_a\partial_y\ep_b
        +\partial_x\ep_b\partial_y\ep_a)}
        {2(i\om_n-\mu-\ep_a)^2(i\om_n-\mu-\ep_b)^2}
        \right]
        g_{xy}^{ab}
        \nonumber\\
        &=
        \frac{e^2}{\hbar^2 c^2}
        \sum_\kv\sum_{a,b}
        g_{xy}^{ab}
        \left[
        2\partial_x\ep_a\partial_y\ep_a
        \lb\frac{f_{ab}}{\ep_{ab}}
        -f'_a + \frac{1}{2}\ep_{ab}f''_a\rb
        -\ep_{ab}\partial_x\partial_y\ep_a
        \lb\frac{f_{ab}}{\ep_{ab}}-f'_a\rb
        +\frac{1}{2}
        (\partial_x\ep_a\partial_y\ep_b
        +\partial_y\ep_a\partial_x\ep_b)
        \lb2\frac{f_{ab}}{\ep_{ab}}
        -f'_a-f'_b\rb
        \right]
        \nonumber\\
        &=
        \frac{e^2}{\hbar^2 c^2}
        \sum_\kv\sum_{a,b}
        g_{xy}^{ab}
        \left[
        (2\partial_x\ep_a\partial_y\ep_a
        -\ep_{ab}\partial_x\partial_y\ep_a
        +\partial_x\ep_a\partial_y\ep_b
        +\partial_y\ep_a\partial_x\ep_b)
        \lb\frac{f_{ab}}{\ep_{ab}}
        -f'_a\rb
        + \ep_{ab}\partial_x\ep_a\partial_y\ep_af''_a
        \right]\;.
    \end{align}
    Similarly, $\tilde\chi_6$ and $\tilde\chi_7$ can be shown to depend on both $g_{xx}^{ab}$ and $g_{yy}^{ab}$, and are equal to each other
    \begin{align}
        \tilde\chi_6
        &=
        \frac{e^2}{2\hbar^2 c^2}k_BT
        \sum_n\sum_\kv\sum_{a,b,c,d}
        \frac{\ep_b\partial_y\ep_a\partial_y\ep_d}
        {(i\om_n+\mu-\ep_a)^2(i\om_n+\mu-\ep_c)}
        \Tr[P_a\partial_xP_bP_c\partial_xP_d]
        +(x\leftrightarrow y)
        \nonumber\\
        &=
        \frac{e^2}{2\hbar^2 c^2}k_BT
        \sum_n\sum_\kv\sum_{a,b,c,d}
        \frac{\ep_b\partial_y\ep_a\partial_y\ep_d}
        {(i\om_n+\mu-\ep_a)^2(i\om_n+\mu-\ep_c)}
        \xi^{ac}_x\xi^{ca}_x
        (\d_{bc}-\d_{ab})(\d_{cd}-\d_{ad})
        +(x\leftrightarrow y)
        \nonumber\\
        &=
        \frac{e^2}{2\hbar^2 c^2}k_BT
        \sum_n\sum_\kv\sum_{a,b}
        g_{xx}^{ab}
        \frac{\ep_{ab}\partial_y\ep_a(\partial_y\ep_a-\partial_y\ep_b)}
        {(i\om_n+\mu-\ep_a)^2(i\om_n+\mu-\ep_b)}
        +(x\leftrightarrow y)\;,
        \\
        \tilde\chi_7
        &=
        \frac{e^2}{2\hbar^2 c^2}k_BT
        \sum_n\sum_\kv\sum_{a,b,c,d}
        \frac{\partial_y\ep_b\partial_y\ep_d}
        {(i\om_n+\mu-\ep_a)(i\om_n+\mu-\ep_c)}
        \Tr[\partial_xP_aP_bP_c\partial_xP_d]
        +(x\leftrightarrow y)
        \nonumber\\
        &=
        \frac{e^2}{2\hbar^2 c^2}k_BT
        \sum_n\sum_\kv\sum_{a,b,c,d}
        \frac{\partial_y\ep_c\partial_y\ep_a}
        {(i\om_n+\mu-\ep_b)(i\om_n+\mu-\ep_d)}
        \Tr[\partial_xP_a\partial_xP_bP_cP_d]
        +(x\leftrightarrow y)
        \nonumber\\
        &=
        \frac{e^2}{2\hbar^2 c^2}k_BT
        \sum_n\sum_\kv\sum_{a,b,c,d}
        \frac{\partial_y\ep_c\partial_y\ep_a}
        {(i\om_n+\mu-\ep_b)(i\om_n+\mu-\ep_c)}
        \lb-\xi_x^{ba}\xi_x^{ab}\d_{bc}
        +\xi_x^{ca}\xi_x^{ac}\d_{ab}
        +\braket{\partial_x a}{\partial_x a}\d_{bc}\d_{ac}
        -\xi_x^{ab}\xi_x^{ba}\d_{ac}
        \rb
        + (x\leftrightarrow y)
        \nonumber\\
        &=
        \frac{e^2}{2\hbar^2 c^2}k_BT
        \sum_n\sum_\kv\sum_{a,b}
        \xi_x^{ab}\xi_x^{ba}
        \frac{\partial_y\ep_a(\partial_y\ep_a-\partial_y\ep_b)}
        {i\om_n+\mu-\ep_a}
        \lb\frac{1}{i\om_n+\mu-\ep_a}
        -\frac{1}{i\om_n+\mu-\ep_b}\rb
        + (x\leftrightarrow y)
        \nonumber\\
        &=
        \frac{e^2}{2\hbar^2 c^2}k_BT
        \sum_n\sum_\kv\sum_{a,b}
        \frac{\ep_{ab}\partial_y\ep_a(\partial_y\ep_a-\partial_y\ep_b)}
        {(i\om_n+\mu-\ep_a)^2(i\om_n+\mu-\ep_b)}
        g_{xx}^{ab}
        + (x\leftrightarrow y)
        \nonumber\\
        &=\tilde\chi_6\;,
    \end{align}
    where we have used Eqs.~(\ref{PmuPnu},\;\ref{PmuPnu_2}). Thus, analogously to $\chi_{xx}$ and $\chi_{yy}$, we have $\tilde{\chi}_{xx}$ and $\tilde{\chi}_{yy}$
    \begin{align}
        \tilde\chi_{xx}
        &=
        \frac{e^2}{\hbar^2 c^2}k_BT
        \sum_n\sum_\kv\sum_{a,b}
        \frac{\ep_{ab}\partial_y\ep_a(\partial_y\ep_a-\partial_y\ep_b)}
        {(i\om_n+\mu-\ep_a)^2(i\om_n+\mu-\ep_b)}
        g_{xx}^{ab}
        \nonumber\\
        &=
        \frac{e^2}{\hbar^2 c^2}
        \sum_\kv\sum_{a,b}
        \partial_y\ep_a(\partial_y\ep_a-\partial_y\ep_b) g_{xx}^{ab}
        \lb
        f'_a - \frac{f_{ab}}{\ep_{ab}}
        \rb\;,
        \\
        \tilde{\chi}_{yy}
        &=
        \frac{e^2}{\hbar^2 c^2}
        \sum_\kv\sum_{a,b}
        \partial_x\ep_a(\partial_x\ep_a-\partial_x\ep_b) g_{yy}^{ab}
        \lb
        f'_a - \frac{f_{ab}}{\ep_{ab}}
        \rb\;.
    \end{align}
    Combining with $\chi_{xx}$ and $\chi_{yy}$, we get
    \begin{align}
        \chi_{xx} + \tilde\chi_{xx}
        &=
        \frac{e^2}{\hbar^2 c^2}
        \sum_\kv\sum_{a,b}
        \partial_y\ep_a(\partial_y\ep_a+\partial_y\ep_b) g_{xx}^{ab}
        \lb
        f'_a - \frac{f_{ab}}{\ep_{ab}}
        \rb\;,\nonumber\\
        \chi_{yy} + \tilde\chi_{yy}
        &=
        \frac{e^2}{\hbar^2 c^2}
        \sum_\kv\sum_{a,b}
        \partial_x\ep_a(\partial_x\ep_a+\partial_x\ep_b) g_{yy}^{ab}
        \lb
        f'_a - \frac{f_{ab}}{\ep_{ab}}
        \rb\;.
    \end{align}
   
    For $\tilde\chi_8,\;\tilde\chi_9\cdots \tilde\chi_{11}$, these terms can be identified as three-band contributions, thus are not of direct importance for a pair of Euler bands. For $\tilde\chi_{12}$, we have a two-band contribution:
    \begin{align}
        \tilde\chi_{12}
        &=
        \frac{e^2}{2\hbar^2 c^2}
        \sum_n
        \sum_\kv\sum_{a,b}
        \frac{\ep_{ab}^2
        \braket{a}{\partial_x b}
        \braket{\partial_y b}{a}}
        {(i\om_n+\mu-\ep_a)^2
        (i\om_n+\mu-\ep_b)}
        \left[\ep_b\lb
        \braket{\partial_x a}
        {b}
        \braket{b}
        {\partial_y a}
        +
        \braket{\partial_y a}
        {b}
        \braket{b}
        {\partial_x a}
        \rb
        +
        \ep_a\lb
        \braket{a}
        {\partial_x\partial_y a}
        +
        \braket
        {\partial_x\partial_y a}
        {a}
        \rb
        \right]
        +
        (x\leftrightarrow y)
        \nonumber\\
        &=
        \frac{e^2}{2\hbar^2 c^2}
        \sum_n
        \sum_\kv\sum_{a,b}
        \frac{\ep_{ab}^2
        \braket{a}{\partial_x b}
        \braket{\partial_y b}{a}}
        {(i\om_n+\mu-\ep_a)^2
        (i\om_n+\mu-\ep_b)}
        \ep_{ba}\lb
        \braket{\partial_x a}
        {b}
        \braket{b}
        {\partial_y a}
        +
        \braket{\partial_y a}
        {b}
        \braket{b}
        {\partial_x a}
        \rb
        +
        (x\leftrightarrow y)
        \nonumber\\
        &=
        2\frac{e^2}{2\hbar^2 c^2}
        \sum_n\sum_\kv\sum_{a,b}
        \ep_{ab}^2(g^{ab}_{xy})^2
        \left[
        -\frac{1}
        {(i\om_n+\mu-\ep_a)^2}
        +
        \frac{1}
        {(i\om_n+\mu-\ep_a)
        (i\om_n+\mu-\ep_b)}
        \right]
        \nonumber\\
        &=
        -2\frac{e^2}{2\hbar^2 c^2}\sum_\kv\sum_{a,b}
        \ep_{ab}^2(g^{ab}_{xy})^2
        \lb f_a'
        -
        \frac{f_{ab}}{\ep_{ab}}
        \rb\; ,
    \end{align}
    which is equal and opposite to $\chi^{(2)}_\mathrm{geo}$ in two-band limit [Eq.~\eqref{chigeo2_2}], given the Hamiltonian is real (i.e., for Euler topology). Thus, these sum up to zero, and will be neglected in the following discussion concerning Euler bands.
    \subsection{Application to Euler bands}
For a pair of centrosymmetric Euler bands $a,b$ with nodal Euler class $e_2$, as mentioned in the main text ansatz, the multiband geometry can be characterized by metric elements~\cite{Jankowski2025PRBoptical}:
\begin{align}
    g^{ab}_{xx} &= \frac{|e_2|^2 k_y^2}{(k_x^2+k_y^2)^2}
    \ ,
    \\
    g^{ab}_{yy} &= \frac{|e_2|^2 k_x^2}{(k_x^2+k_y^2)^2}
    \ ,
    \\
    g^{ab}_{xy} &= -\frac{|e_2|^2 k_xk_y}{(k_x^2+k_y^2)^2}
    \ ,
\end{align}
while the Berry curvature $\Om^{ab}_{xy}$ vanishes under the $\mathcal{PT}$ symmetry. With $\ep_i = {k^2}/{2m_i}$, where $i=a,\ b$, we have $\chi_\text{E}$ which is the same as in Eq.~\eqref{chiE0}. 
For the remaining contributions, one has to consider the cases: $\sgn(m_a)=\sgn(m_b)$ and $\sgn(m_a)=-\sgn(m_b)$ separately, which, for instance, correspond to distinct band structures around the Euler nodes in Lieb lattice cuprates and graphene bilayers, respectively.
We are particularly interested in the case of $\sgn(m_a)=\sgn(m_b)$.
We begin by noting the following in the limit of $T\ra 0$ and $\mu\neq 0$:
\begin{align}
    \sum_\kv \frac{f_{ab}}{k^2}g(\th)
    &=
    \frac{1}{4\pi^2}
    \int^{2\pi}_0d\th\; g(\th)
    \int^\infty_0 kdk\frac{f_{ab}}{k^2}
    \nonumber\\
    &\ra
    \frac{\sgn(m_a)}{4\pi^2}
    \int^{2\pi}_0d\th\; g(\th)
    \int^{\sqrt{2m_a\mu}}_{\sqrt{2m_b\mu}}
    \frac{dk}{k}
    \nonumber\\
    &=
    \frac{\sgn(m_a)}{8\pi^2}
    \ln{\frac{m_a}{m_b}}
    \int^{2\pi}_0d\th\; g(\th)
    \ ,
    \\
    \sum_\kv {f'_a}g(\th)
    &\ra
    -\frac{1}{4\pi^2}
    \int^{2\pi}_0d\th\; g(\th)
    \int^\infty_0 kdk\; \d(\ep_a-\mu)
    \nonumber\\
    &=
    -\frac{1}{4\pi^2}
    \int^{2\pi}_0d\th\; g(\th)
    \int^\infty_{-\infty} |m_a|d\ep_a\; \d(\ep_a-\mu)
    \nonumber\\
    &=
    -\frac{|m_a|}{4\pi^2}
    \int^{2\pi}_0d\th\; g(\th)
    \ ,
    \\
    \sum_\kv f''_ag(\th)
    &\ra 0,
\end{align}
which are the only relevant sums over $k$-space. Using $k_x = k\cos\th$ and $k_y = k\sin\th$, we thus get:
\begin{align}
    \chi_\text{LP}
    &=
    \frac{e^2}{6\hbar^2 c^2}\sum_{i=a,b}\sum_\kv
    \left[
    \frac{\partial^2\ep_i}
    {\partial k_x^2}
    \frac{\partial^2\ep_i}
    {\partial k_y^2}
    -
    \lb
    \frac{\partial^2\ep_i}
    {\partial k_x\partial k_y}
    \rb^2
    \right]
    f'_i
    \nonumber\\
    &\ra
    -\frac{e^2}{6\hbar^2 c^2}
    \sum_{i=a,b}\sum_\kv
    \frac{1}{m_i^2}\d(\ep_i-\mu)
    \nonumber\\
    &=
    -
    \frac{e^2}{12\pi\hbar^2 c^2}
    \sum_{i=a,b}
    \frac{1}{|m_i|}
    \ , 
    \\
    \chi_{xy} + \tilde\chi_{xy}
    &\ra
    \frac{e^2}{\hbar^2 c^2}
    \sum_\kv
    g_{xy}^{ab}
    (2\partial_x\ep_a\partial_y\ep_a
    -\ep_{ab}\partial_x\partial_y\ep_a
    +\partial_x\ep_a\partial_y\ep_b
    +\partial_y\ep_a\partial_x\ep_b)
    \lb\frac{f_{ab}}{\ep_{ab}}
    -f'_a\rb
    + (a\leftrightarrow b)
    \nonumber\\
    &=
    -2\frac{e^2}{\hbar^2 c^2}\sum_\kv 
    \sin^2\th\cos^2\th\left[
    \frac{f_{ab}}{k^2}
    \frac{(m_a+m_b)^2}{m_am_b(m_b-m_a)}
    -
    \frac{(m_a+m_b)f_{a}'}{m_a^2m_b}
    -
    \frac{(m_a+m_b)f_{b}'}{m_b^2m_a}
    \right]
    \nonumber\\
    &\ra
    -2\frac{e^2}{\hbar^2 c^2}\frac{\pi}{4}
    \sgn(m_a)\left[
    -\frac{2(m_a+m_b)^2}{m_am_b(m_a-m_b)}
    \frac{1}{8\pi^2}
    \ln\frac{m_a}{m_b}
    +
    \frac{1}{2\pi^2}
    \frac{m_a+m_b}{m_a m_b}
    \right]
    \nonumber\\
    &=
    \frac{e^2}{4\pi\hbar^2 c^2}
    \sgn(m_a)\left[
    \frac{(m_a+m_b)^2}{2m_am_b(m_a-m_b)}
    \ln\frac{m_a}{m_b}
    -
    \frac{m_a+m_b}{m_a m_b}
    \right]
    \ ,
    \\
    \chi_{xx} + \tilde\chi_{xx}
    &=
    \frac{e^2}{\hbar^2 c^2}
    \sum_\kv
    \partial_y\ep_a(\partial_y\ep_a+\partial_y\ep_b) g_{xx}^{ab}
    \lb
    f'_a - \frac{f_{ab}}{\ep_{ab}}
    \rb
    +(a\leftrightarrow b)
    \nonumber\\
    &=
    \frac{e^2}{\hbar^2c^2}\sum_\kv 
    \sin^4\th
    \left[
    \lb\frac{1}{m_a}+\frac{1}{m_b}\rb
    \lb\frac{f'_a}{m_a}
    +
    \frac{f'_b}{m_b}\rb
    -
    \frac{2(m_a+m_b)^2}{m_am_b(m_b-m_a)}
    \frac{f_{ab}}{k^2}
    \right]
    \nonumber\\
    &\ra
    -\frac{e^2}{\hbar^2c^2}
    \frac{3\pi}{4}
    \sgn(m_a)
    \left[
    \frac{1}{2\pi^2}
    \frac{m_a+m_b}{m_a m_b}
    -
    \frac{2(m_a+m_b)^2}{m_am_b(m_b-m_a)}
    \frac{1}{8\pi^2}
    \ln\frac{m_a}{m_b}
    \right]
    \nonumber\\
    &=
    \frac{3e^2}{8\pi\hbar^2 c^2}
    \sgn(m_a)\left[
    \frac{(m_a+m_b)^2}{2m_am_b(m_a-m_b)}
    \ln\frac{m_a}{m_b}
    -
    \frac{m_a+m_b}{m_a m_b}
    \right]
    \ ,
    \\
    \chi_{yy}
    + \tilde\chi_{yy}
    &=
    \chi_{xx}
    + \tilde\chi_{xx}
    =
    \frac{3}{2}(\chi_{xy}+\tilde\chi_{xy})
    \ .
\end{align}
As such, when $m_a\ra m_b$, only the energetic term has a nonvanishing contribution. 
We note that the magnetic susceptibility is in general constant in value around the node, as we have predicted from magnetic response function using Landau level, provided the bands are perfectly quadratic, and at zero temperature.
Additionally, the geometrical contribution is always opposite to the energetic contribution for this ideal case.

\subsubsection*{Higher angular harmonic contributions}
For realistic models, where the full rotational symmetry is broken, one may instead only have $\mathcal{C}_4\mathcal{T}$ symmetry instead of the $\mathcal{PT}$ symmetry. Below, we discuss the contribution due to the higher angular harmonic terms in both the band dispersion and geometry. We begin with the definition of patch Euler class, using an infinitesimally small contour around the node:
\begin{align}
    e_2 &= \frac{1}{2\pi}\oint_{|\kv|\ra 0} d\kv \cdot \boldsymbol{\xi}
    \nonumber\\
    &= \frac{k}{2\pi} \int_0^{2\pi} d\th (-\xi_x\sin\th+\xi_y\cos\th).
\end{align}
Given that one can perform a Fourier transform of $\xi_\mu$ as a function of $\th$, and assume $\xi_x(\th+\pi/2)=\xi_y(\th)$, one must have:
\begin{align}
    \xi_\mu(k,\th)= \frac{1}{k} [e_2 (-\sin\th, \cos\th) + \tilde{\xi}_\mu(\th)],
\end{align}
where $\tilde{\xi}_\mu$ is the higher periodicity contribution -- in our case, most likely of form $\tilde{\xi}\propto \sin 3\th$. As such, the patch Euler class does not fully constrain the band geometry at the higher harmonic orders. \\

As for the band dispersion in the considered model, the higher angular periodicity terms can be also observed. In particular, the dispersion of one of the bands can be approximated as:
\begin{equation}
    \ep(k,\th) \sim k^2(1-\a \cos 4\th),
\end{equation}
where $\a\sim 1$ captures the relative strength of the higher angular harmonic.

\section{Details on Matsubara sums}\label{app::C}
In this section, we compute the required Matsubara sums. For an odd Matsubara sum  $k_BT\sum_n g(i\om_n)$, we evaluate
\begin{align}
    k_BT\sum_n g(i\om_n)
    &=
    -\frac{1}{2i\pi}\oint dz\ g(z)f(z)
    \nonumber\\
    &=
    \sum_{z_0} \mathrm{Res}_{z=z_0}[g(z)f(z)]\ ,
\end{align}
where $z_0$ are poles of the function $g(z)$. We begin by noting that for $g(i\om)=(i\om+\mu-\ep_a)^{-n}$ with pole $z_0 = \ep_a-\mu$:
\begin{align}
    k_BT\sum_n (i\om_n+\mu-\ep_a)^{-n}
    &=
    \mathrm{Res}_{z=\ep_a-\mu}[(z+\mu-\ep_a)^{-n}f(z)]
    \nonumber\\
    &=
    \frac{1}{(n-1)!}\mathrm{Res}_{z=\ep_a-\mu}[(z+\mu-\ep_a)^{-1}\partial_z ^{n-1}f(z)]
    \nonumber\\
    &=
    \frac{1}{(n-1)!}f^{(n)}_a\label{MS}\ ,
\end{align}
where $f^{(n)}_a=\partial_z ^{n-1}f(z)|_{z=\ep_a-\mu}$. As such, for $\ep_a\neq \ep_b$:
\begin{align}
    k_BT\sum_n \frac{1}{(i\om_n+\mu-\ep_a)^2(i\om_n+\mu-\ep_b)^2}
    &=
    \partial_z\left.\frac{f(z)}{(z+\mu-\ep_b)^2}\right|_{z=\ep_a-\mu}
    +
    (a\leftrightarrow b)
    \nonumber\\
    &=
    \ep_{ab}^{-2}f'_a - 2\ep_{ab}^{-3}f_a + (a\leftrightarrow b)
    \nonumber\\
    &=
    \frac{1}{\ep_{ab}^{2}}\lb
    f'_a+f'_b
    -
    2\frac{f_{ab}}{\ep_{ab}}
    \rb\ , \label{MS_22}
    \\
    k_BT\sum_n \frac{1}{(i\om_n+\mu-\ep_a)^2(i\om_n+\mu-\ep_b)}
    &=
    \partial_z\left.\frac{f(z)}{z+\mu-\ep_b}\right|_{z=\ep_a-\mu}
    +
    \left.\frac{f(z)}{(z+\mu-\ep_a)^2}\right|_{z=\ep_b-\mu}
    \nonumber\\
    &=
    \ep_{ab}^{-1}f'_a-\ep_{ab}^{-2}f_a+\ep_{ab}^{-2}f_b
    \nonumber\\
    &= \frac{1}{\ep_{ab}}\lb
    f'_a - \frac{f_{ab}}{\ep_{ab}}
    \rb
    \ ,
    \label{MS_21}
    \\
    k_BT\sum_n \frac{1}{(i\om_n+\mu-\ep_a)^3(i\om_n+\mu-\ep_b)}
    &=
    \frac{1}{2!}\partial_z^2\left.\frac{f(z)}{z+\mu-\ep_b}\right|_{z=\ep_a-\mu} + \left.\frac{f(z)}{(z+\mu-\ep_a)^3}\right|_{z=\ep_b-\mu}
    \nonumber\\
    &=
    \frac{f_a}{\ep_{ab}^3}  
    -
    \frac{f_a'}{\ep_{ab}^2}
    +
    \frac{f_a''}{2\ep_{ab}}
    +
    \frac{f_b}{\ep_{ba}^3}
    \nonumber\\
    &=
    \frac{f_{ab}}{\ep_{ab}^3}  
    -
    \frac{f_a'}{\ep_{ab}^2}
    +
    \frac{f_a''}{2\ep_{ab}}
    \ ,\label{MS_31}
\end{align}
\begin{align}
    &k_BT\sum_n \frac{1}{(i\om_n+\mu-\ep_a)^2(i\om_n+\mu-\ep_b)(i\om_n+\mu-\ep_c)}
    \nonumber\\
    &=
    \partial_z\left.\frac{f(z)}{(z+\mu-\ep_b)(z+\mu-\ep_c)}\right|_{z=\ep_a-\mu}
    +
    \left.\frac{f(z)}{(z+\mu-\ep_a)^2(z+\mu-\ep_c)}\right|_{z=\ep_b-\mu}
    +
    \left.\frac{f(z)}{(z+\mu-\ep_a)^2(z+\mu-\ep_b)}\right|_{z=\ep_c-\mu}
    \nonumber\\
    &=
    \frac{f'_a}{\ep_{ab}\ep_{ac}}
    -
    \frac{f_a}{\ep_{ab}^2\ep_{ac}}
    -
    \frac{f_a}{\ep_{ab}\ep_{ac}^2}
    +
    \frac{f_b}{\ep_{ba}^2\ep_{bc}}
    +
    \frac{f_c}{\ep_{ca}^2\ep_{cb}}
    \ ,
    \label{MS_211}
    \\
    &k_BT\sum_n \frac{1}{(i\om_n+\mu-\ep_a)(i\om_n+\mu-\ep_b)(i\om_n+\mu-\ep_c)(i\om_n+\mu-\ep_d)}
    \nonumber\\
    &=\frac{f_a}{\ep_{ab}\ep_{ac}\ep_{ad}}
    +
    \frac{f_b}{\ep_{ba}\ep_{bc}\ep_{bd}}
    +
    \frac{f_c}{\ep_{ca}\ep_{cb}\ep_{cd}}
    +
    \frac{f_d}{\ep_{da}\ep_{db}\ep_{dc}}
    \ ,
\end{align}
where we have defined $f_{ab}= f_a-f_b$. We next use integration by parts to simplify the higher-order derivative $f^{(n)}_a$ to $f_a'$, which approaches $-\d(\ep_a-\mu)$ as the temperature approaches zero. Specifically,
\begin{align}
    \sum_\kv g(\kv)f''_a
    &=
    \sum_\kv g(\kv)\lb\frac{\partial\ep_a}{\partial k_\nu}\rb^{-1}\frac{\partial}{\partial k_{\nu}} f'_a
    \nonumber\\
    &=
    \sum_\kv \frac{\partial}{\partial k_{\nu}}\left[g(\kv)\lb\frac{\partial\ep_a}{\partial k_\nu}\rb^{-1} f'_a\right]
    -
    \frac{\partial}{\partial k_{\nu}}\left[g(\kv)\lb\frac{\partial\ep_a}{\partial k_\nu}\rb^{-1} \right] f'_a
    \nonumber\\
    &\ra
    \sum_\kv
    \frac{\partial}{\partial k_{\nu}}\left[g(\kv)\lb\frac{\partial\ep_a}{\partial k_\nu}\rb^{-1} \right] 
    \d(\ep_a-\mu), \label{fpp}
    \\
    \sum_\kv g(\kv)f'''_a
    &=
    \sum_\kv g(\kv)\lb\frac{\partial\ep_a}{\partial k_{\nu_1}}\rb^{-1}\frac{\partial}{\partial k_{\nu_1}} f''_a
    \nonumber\\
    &\ra
    -\sum_\kv
    \frac{\partial}{\partial k_{\nu_1}}\left[g(\kv)\lb\frac{\partial\ep_a}{\partial k_{\nu_1}}\rb^{-1} \right] 
    f''_a
    \nonumber\\
    &\ra
    -
    \sum_\kv
    \frac{\partial}{\partial k_{\nu_2}}\left[
    \lb\frac{\partial\ep_a}{\partial k_{\nu_2}}\rb^{-1} 
    \frac{\partial}{\partial k_{\nu_1}}\left[g(\kv)
    \lb\frac{\partial\ep_a}{\partial k_{\nu_1}}\rb^{-1} 
    \right] 
    \right]
    \d(\ep_a-\mu),
\end{align}
where $\nu_i$ can be $x$ or $y$. This identity is then used for rewriting the energetic contribution $\chi_\text{E}$ obtained in Eq.~\eqref{chiE}.

\end{widetext}

\bibliography{main}
\end{document}